\documentclass[a4paper,11pt]{article}
\usepackage{jheppub} % for details on the use of the package, please see the JINST-author-manual
\usepackage{amsmath}
\usepackage{amssymb}
\usepackage{lineno}
\usepackage{tikz}
\usepackage{mathrsfs}
\usepackage{placeins}
\usepackage{float}
\usetikzlibrary{arrows.meta, positioning}
\renewcommand{\=}{\;=\;}

\title{\boldmath Geometry and holography of $\mathrm{AdS}_2\times\mathrm{S}^2$:
towards the field theory dual%Two non-Abelian T-dual routes to $\mathrm{AdS}_2 \times\mathrm{S}^2$: geometry and dual quantum mechanics
}

\author{Miroslava Mosso Rojas$^a$,}
\affiliation{$^a$ PRISMA+ Cluster of Excellence, Mainz Institute for Theoretical Physics,
\\Johannes Gutenberg-Universität Mainz, D - 55099 Mainz, Germany}

\author{Anayeli Ramirez$^b$}
\affiliation{$^b$ Instituut voor Theoretische Fysica, KU Leuven,\\
Celestijnenlaan 200D, 3000 Leuven, Belgium}

\emailAdd{mmossoro@uni-mainz.de, mariaanayeli.ramirezortiz@kuleuven.be}

\abstract{We take a first step in identifying the superconformal quantum mechanics dual to supersymmetric AdS$_2\times$S$^2$ solutions. We focus on the Type IIB solutions of the form AdS$_2\times$S$^2\times$CY$_2\times\Sigma$ that preserve eight supercharges, with $\Sigma$ a two-dimensional space parametrised by $(x_1,x_2)$, and read off their brane content, a D1--D5--NS5 intersection with D3 and D7 flavour branes. We propose that their dual $\mathcal{N}=4$ quantum mechanics is organised over a two-dimensional lattice of cells in $\Sigma$. Namely, it consists of copies of the linear quiver along $x_1$ dual to the U(1)-invariant solutions of the class, one for each interval along $x_2$, with ranks given by the D1 and D5 charges of the corresponding cell. Its central charge agrees with the holographic one at leading order. We show that the family of SL$(2,\mathbb{R})$ non-Abelian T-duals of the AdS$_3\times$S$^2\times$CY$_2$ solutions of massive Type IIA supergravity, constructed in previous work, belongs to this class, as does the SL$(2,\mathbb{R})\times$SU$(2)$ non-Abelian T-dual of AdS$_3\times$S$^3\times$CY$_2$, and work both out in detail. For both we propose global completions that introduce additional D3 and D7 branes. Along $x_2$ we interpret the fundamental strings stretched between the colour branes and the additional D3 and D7 branes as baryon vertices. Finally, we extend this proposal from the non-Abelian T-duals to the solutions of the class whose defining functions factorise in $x_1$ and $x_2$, including those in which the AdS$_2$ and S$^2$ factors have the same warp factor.
}

\preprint{MITP-26-052}

\begin{document}

\maketitle

\flushbottom

\section{Introduction}
The Bekenstein--Hawking formula relates the entropy of a black hole to the area of its horizon, and suggests that the geometry encodes a large number of microscopic degrees of freedom. Identifying these degrees of freedom and explaining their degeneracy from first principles is one of the central problems of quantum gravity. For a class of supersymmetric black holes in string theory this was achieved by counting bound states of D-branes \cite{Strominger:1996sh}. Extremal black holes are especially well suited to this programme. Their Hawking temperature vanishes, their near-horizon geometry has enhanced symmetry and, in supersymmetric settings, many of their properties are protected against quantum corrections. Moreover, quantities such as the entropy can be computed directly from the near-horizon geometry \cite{Sen:2005wa,Sen:2008vm}.

Supersymmetric extremal black holes typically develop an infinitely deep throat near the horizon, whose geometry contains an AdS$_{2}$ factor. In the spirit of the AdS/CFT correspondence, a supersymmetric AdS$_{2}$ solution should then be dual to a superconformal quantum mechanics that describes the infrared dynamics of the corresponding brane system. Such a description would provide, in principle, a microscopic framework in which the entropy and other protected observables of the black hole can be computed. However, the AdS$_{2}/$CFT$_{1}$ correspondence is much less understood than its higher-dimensional counterparts. The boundary of AdS$_{2}$ is disconnected, finite-energy excitations backreact strongly on the geometry, and it is not clear how to separate the degrees of freedom of the throat from those of its ultraviolet completion \cite{Strominger:1998yg,Maldacena:1998uz,Sen:2008vm}. It is therefore desirable to construct explicit AdS$_{2}$ solutions of ten- and eleven-dimensional supergravity and, whenever possible, to identify their dual quantum mechanics.
\FloatBarrier
\label{sec:intro}
\begin{figure}[t]
    \centering
    \fbox{
        \parbox{0.80\linewidth}{
            \centering
            \includegraphics[width=0.85\textwidth]{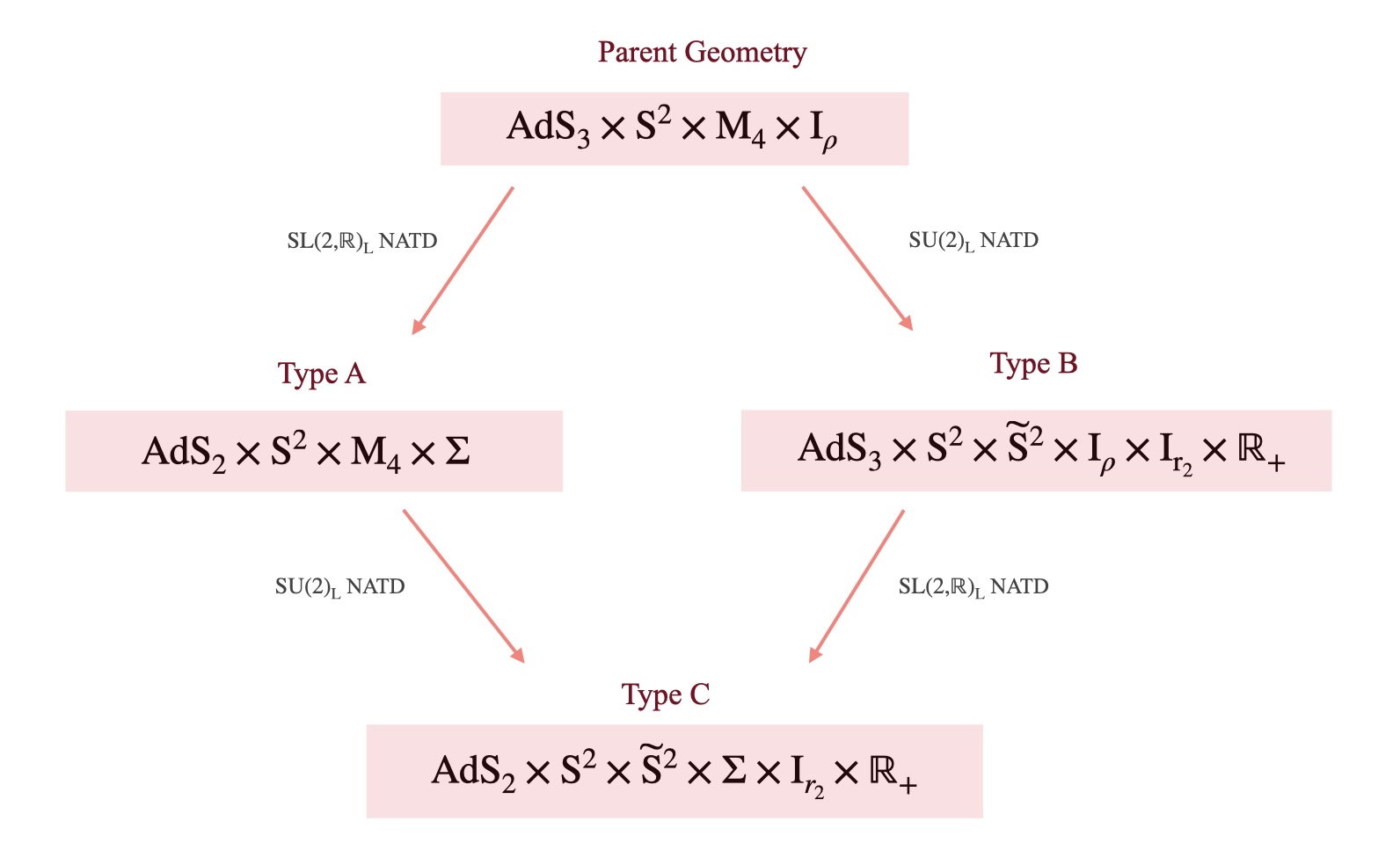}
        }
    }
    \caption{
    Schematic representation of the two NATD chains, which start from the same parent geometry and end on the same Type C geometry. $\mathrm{M}_{4}$ denotes a real four-dimensional manifold. The Type A geometry can be obtained for $\mathrm{M}_{4}=\mathrm{CY_2}$ or $\mathrm{M}_{4}=\mathrm{S}^3 \times \mathbb{R}$, while the Type B and Type C geometries require $\mathrm{M}_{4}=\mathrm{S}^3 \times \mathbb{R}$. $\Sigma$ is a two-dimensional real manifold.
    }
    \label{Fig::Diagram}
\end{figure}

Many supersymmetric AdS$_{2}$ backgrounds have been studied in recent years. They include near-horizon geometries of  intersecting branes, reductions and T-duals of higher-dimensional AdS solutions, compactifications of holographic field theories on Riemann surfaces or higher-dimensional manifolds, and geometries describing conformal line defects in higher-dimensional CFTs \cite{DHoker:2007mci, Chiodaroli:2009yw, Chiodaroli:2009xh,Dibitetto:2018gbk,Gutperle:2018fea,Dibitetto:2018gtk,Chen:2019qib,
Hong:2019wyi, Chen:2020mtv,Dibitetto:2019nyz, Lozano:2020txg,Lozano:2020sae,Lozano:2021rmk,
Ramirez:2021tkd,Lozano:2021xxs,Lozano:2021fkk,Lozano:2022vsv,Lozano:2022swp,Conti:2023naw,Conti:2023rul,Karndumri:2024gtv,Legramandi:2023fjr,Conti:2025djz,Lozano:2025ief,Kim:2006qu,Gauntlett:2006ns,Macpherson:2025aau,Conti:2025wyj,Conti:2025wwf}. Necessary and sufficient conditions for supersymmetric AdS$_{2}$ solutions of Type II and eleven-dimensional supergravity have  been derived using spinor bilinears and $G$-structure techniques \cite{Kim:2006qu,Gauntlett:2006ns,Hong:2019wyi,Legramandi:2023fjr,Macpherson:2025aau}. The space of solutions they describe is very large, as one may expect from the high dimension of the internal space. Among them, Type IIB backgrounds with warped AdS$_{2}\times$S$^{2}$ factors \cite{DHoker:2007mci,Chiodaroli:2009yw,Chiodaroli:2009xh,Lozano:2020txg,Lozano:2021rmk,Ramirez:2021tkd,Lozano:2021fkk} are of special interest, since AdS$_{2}\times$S$^{2}$ is the near-horizon geometry of static, charged extremal black holes in four-dimensional asymptotically flat spacetimes. Still, relatively few AdS$_{2}/$CFT$_{1}$ pairs are understood on both sides of the correspondence, and new controlled examples are needed.

Non-Abelian T-duality (NATD) provides a systematic route towards such solutions. It was formulated as a generalisation of the Buscher procedure to non-Abelian isometry groups \cite{delaOssa:1992vci}, and it became a useful solution-generating technique in supergravity once the transformation of the Ramond--Ramond sector was established \cite{Sfetsos_2011}. 
Unlike Abelian T-duality, NATD is not known to be an exact symmetry of string theory, and the local transformation does not fix the global properties of the dual coordinates \cite{Sfetsos_2011,Giveon:1993ai,Alvarez:1993qi,Alvarez:1994np}. Supplemented with flux quantisation, large gauge transformations and an analysis of the brane sources, it has nevertheless generated many holographic backgrounds \cite{Itsios:2012dc,Lozano:2012au,Itsios:2013wd,Barranco:2013fza,Gaillard:2013vsa,Caceres:2014uoa,Macpherson:2014eza,Kelekci:2014ima,Lozano:2014ata,Lozano:2015bra,Lozano:2015cra,Macpherson:2015tka,Kooner:2014cqa} and, in several cases, concrete proposals for their dual field theories \cite{Lozano:2013oma,Lozano:2016kum,Lozano:2016wrs,Itsios:2017cew,Lozano:2017ole,Lozano:2018pcp,Lozano:2019ywa,vanGorsel:2017goj,Zacarias:2014wta}. Most applications involve SU$(2)$ isometries acting on an internal three-sphere, see \cite{Thompson:2019ipl} for a review. The dualisation then replaces the S$^{3}$ by a non-compact radial direction and an S$^{2}$, which makes it a natural way to produce backgrounds with two-sphere factors.

More recently, NATD has been extended to freely acting SL$(2,\mathbb{R})$ isometries \cite{Lozano:2021rmk,Ramirez:2021tkd}. The isometry group of AdS$_{3}$ is $\mathrm{SO}(2,2)\simeq\mathrm{SL}(2,\mathbb{R})_{L}\times\mathrm{SL}(2,\mathbb{R})_{R}$, and both factors act freely. Dualising one of them preserves the other, which is realised as the isometry group of an AdS$_{2}$ subspace. SL$(2,\mathbb{R})$ NATD thus replaces AdS$_{3}$ by a warped AdS$_{2}$ factor and an additional non-compact direction, and gives a direct procedure to generate AdS$_{2}$ solutions from known AdS$_{3}$ backgrounds. This construction has already produced new AdS$_{2}$ geometries with proposed quiver quantum mechanics duals, whose interpretation involves brane intersections and baryon vertices \cite{Ramirez:2021tkd,Conti:2025djz,Lozano:2022fqk}. At the level of sigma models, NATD has also been generalised to supergroup and supercoset targets, allowing for dualisation along both bosonic and fermionic directions \cite{Bielli:2021hzg,Bielli:2022gmm}. A related analysis at the level of principal chiral models has shown that, in a suitable limit, the NATD of the SL$(2,\mathbb{R})$ principal chiral model can be recast as a Poisson sigma model describing Jackiw--Teitelboim (JT) gravity \cite{Bielli:2024wru}. %This suggests that the appearance of AdS$_{2}$ under non-compact NATD may be connected with the  gravitational dynamics of near-extremal horizons.

The SL$(2,\mathbb{R})$ and SU$(2)$ dualisations play complementary roles: the first one generates an AdS$_{2}$ factor from AdS$_{3}$, and the second one an S$^{2}$ from the three-sphere. Backgrounds with both AdS$_{3}$ and S$^{3}$ factors are therefore natural starting points to construct AdS$_{2}\times$S$^{2}$ geometries. In this paper we exploit this observation for the massive Type IIA AdS$_{3}\times$S$^{2}\times$CY$_{2}$ solutions of \cite{Lozano:2019emq}. When CY$_{2}=\mathbb{R}^{4}$, written as a cone over S$^{3}$, these solutions have an additional SU$(2)$ isometry acting freely on the S$^{3}$, so that both groups can be dualised. Since the SL$(2,\mathbb{R})$ and SU$(2)$ groups act on independent sectors of the parent background, their Lie algebras form a direct sum and the two dualisations are expected to commute. Dualising first SL$(2,\mathbb{R})$ and then SU$(2)$, or the other way round, gives the same solution, which can also be obtained in a single dualisation with respect to SL$(2,\mathbb{R})\times$SU$(2)$. The two chains and the three resulting geometries are summarised in Figure~\ref{Fig::Diagram}.

We refer to the background obtained by dualising SL$(2,\mathbb{R})$ as Type A. It is a Type IIB solution with an AdS$_{2}\times$S$^{2}$ factor, first constructed in \cite{Lozano:2021rmk}. Dualising the SU$(2)$ instead gives the Type B background, which keeps the AdS$_{3}$ factor and gains a second two-sphere. Dualising both groups gives the Type C background, with an AdS$_{2}$ factor and two two-spheres. While the parent solution has a non-vanishing Romans mass, the Type C background is a solution of massless Type IIA supergravity.

Most of our analysis concerns the holographic interpretation of the Type A backgrounds. We show that they belong to the class of Type IIB AdS$_{2}\times$S$^{2}\times$CY$_{2}\times\Sigma_{2}$ solutions of \cite{Legramandi:2023fjr}, and read off their brane content from the charges. Since NATD only determines the background locally, we propose a global completion in which the solution is glued to its mirror image along the dual direction, as in \cite{Ramirez:2021tkd}, with D3 and D7 branes at the turning point. We propose that the dual quantum mechanics is a collection of copies of the quiver dual to the Abelian T-dual of the parent solution \cite{Lozano:2020txg}, one for each interval along the dual direction. Its central charge agrees with the holographic one at leading order, and fundamental strings stretched between the colour branes and the D3 and D7 branes realise baryon vertices. The double non-Abelian T-dual of AdS$_{3}\times$S$^{3}\times$CY$_{2}$ also belongs to the Type A family. We complete it along both directions, and its dual quantum mechanics takes the same form. Finally, we extend the proposal beyond the Type A backgrounds, to the separable solutions of the class.

The paper is organised as follows. In Section~\ref{sec:NATD} we construct the Type A, B and C backgrounds. In Section~\ref{sec:FT} we embed the Type A backgrounds in the class of \cite{Legramandi:2023fjr} and construct their global completion. Section~\ref{sec:dualQM} contains our proposal for the dual quantum mechanics, and we conclude in Section~\ref{sec:conclusions}. Appendix~\ref{app:groups} collects our conventions for the Maurer--Cartan forms and the single dualisation with respect to SL$(2,\mathbb{R})\times$SU$(2)$, and Appendix~\ref{app:B-Appendix} the parent solution of massive Type IIA.

\section{SL(2,$\mathbb{R}$) and SU(2) non-Abelian T-duality and three dual backgrounds}
\label{sec:NATD}
Non-Abelian T-duality (NATD) provides a systematic method for generating new supergravity backgrounds by extending the Buscher procedure beyond Abelian isometries \cite{delaOssa:1992vci}. We use it as a solution-generating technique \cite{Sfetsos_2011}. Whether it is a symmetry of string theory is not known, although holographic applications suggest that it is not \cite{Lozano:2016kum}. In this work, we consider target spaces admitting a non-Abelian isometry group of the form
\begin{equation}
\mathcal{G} = \mathrm{SL(2,\mathbb{R})} \times \mathrm{SU(2)}\,,
\end{equation}
where each factor acts on independent submanifolds. The corresponding Lie algebra decomposes as a direct sum
\begin{equation}
\mathfrak{g} = \mathfrak{sl}(2,\mathbb{R}) \oplus \mathfrak{su}(2)\,,
\end{equation}
so that generators belonging to different sectors commute. As a consequence, the associated NATD transformations act independently on each factor. In particular, performing the dualisation along the $\mathrm{SL(2,\mathbb{R})}$ directions followed by the $\mathrm{SU(2)}$ ones, or in the reverse order, leads to the same dual background. In the rest of this work we dualise each factor separately.

We begin with the bosonic string sigma model describing the propagation of a string in a background with metric and $B$-field encoded in the tensor $E$. In the presence of a non-Abelian isometry group $\mathcal{G}$. Here  $\mathcal{G}$ denotes either $\mathrm{SL(2,\mathbb{R})}$ or $\mathrm{SU(2)}$ depending on the sector being dualised. The action can be written as
\begin{equation}
    \begin{split}
        S & \= \int d\sigma^{2}  \left( E_{ij} \, L_{+}^{i} L_{-}^{j} + Q_{i \mu} \, L_{+}^{i} \partial_{-} x^{\mu} + Q_{\mu i} \, \partial_{+} x^{\mu} \, L_{-}^{i} + Q_{\mu \nu} \, \partial_{+}x^{\mu} \, \partial_{-}x^{\nu} \right)~, \\
       & \text{with} \quad\quad E_{ij} \= g_{ij} + b_{ij}, \quad\quad Q_{i\mu} \= G_{i \mu} + B_{i \mu},\\
       &\qquad\qquad
        Q_{\mu i} \= G_{\mu i} + B_{\mu i}, \quad\quad Q_{\mu \nu} \= G_{\mu \nu} + B_{\mu \nu}~,
        \end{split}
    \label{actionMC}
\end{equation}
where $i=8,9,10$ label directions along the group manifold and $\mu,\nu = 1,2,...,7$ denote spectator coordinates. 
$G_{\mu \nu}$ is the target space metric and and $B_{\mu \nu}$ the target space NSNS two-form, whereas $g_{ij}$ and $b_{ij}$ are the components along the group directions. Notice that quantities $Q = G + B$ encode the NS sector fields. $L_{\pm}^{i}$ are the pullbacks of the left-invariant Maurer--Cartan one-forms, ensuring invariance of the action under global left multiplication $g \mapsto \lambda g$, with $g$ and $\lambda \in \mathcal{G}$,
\begin{equation}
    L^{i} \= -i \mathrm{Tr}(t^{i} g^{-1} dg)~, \quad \text{such that,} \quad dL^{i} \= \frac{1}{2} f^{i}_{jk} L^{j} \ L^{k}~,
    \label{MCforms}
\end{equation}
with $t^i$ the generators of the Lie algebra $\mathfrak{g}$ associated to $\mathcal{G}$, and $f^i{}_{jk}$ the structure constants. The explicit generators and the algebra they satisfy are collected in Appendix~\ref{app:groups}, see \eqref{eq:algebraSU(2)} for $\mathrm{SU(2)}$ and \eqref{eq:algebraSl(2R)} for $\mathrm{SL(2,\mathbb{R})}$. In both cases, the geometry can be expressed in terms of the corresponding Maurer–Cartan forms, allowing for a unified treatment of compact and non-compact groups, see Appendix \ref{app:groups}.

The non-Abelian dualisation \cite{delaOssa:1992vci} proceeds by gauging the global symmetry associated with the group $\mathcal{G}$. This is achieved by promoting ordinary derivatives to covariant derivatives, $\partial_{\pm} g \, \mapsto \, D_{\pm}g \= \, \partial_{\pm}g - A_{\pm}g~,$ and introducing non-dynamical gauge fields $A_\pm$ valued in the Lie algebra $\mathfrak{g}$. To enforce the flatness of the gauge connection, one adds a Lagrange multiplier term $
- i \, \text{Tr}(v F_{+-})\,$, with $F_{+-} = \partial_+ A_- - \partial_- A_+ - [A_+,A_-]\,$
and $v = (v_1,v_2,v_3)$ is a Lagrange multiplier vector, taking values in the Lie algebra $\mathfrak{g}$ of the group $\mathcal{G}$. 
Therefore, the action (\ref{actionMC}) remains invariant under the local transformations
\begin{equation}
    g \mapsto \lambda^{-1}g~, \quad A_{\pm} \mapsto \lambda^{-1}(A_{\pm} \lambda - \partial_{\pm} \lambda)~, \quad v \mapsto \lambda^{-1} v \lambda~,
\end{equation}
for $\lambda \in \mathcal{G}$. This gauge freedom can be used to fix the group element, and in what follows we adopt the gauge choice $g = \mathbb{I}$.

The dual theory is then obtained by integrating the Lagrange multiplier term by parts and solving for the gauge fields, 
\begin{equation}
    A_{+}^{i} \= i (M^{-1})_{ij} (\partial_{+} v_{j} + Q_{\mu j} \, \partial_{+} x^{\mu})~, \quad A_{-}^{i} \= -i (M^{-1})_{ij}(\partial_{-} v_{j}-Q_{j\mu} \,\partial_{-}x^{\mu})~,
\end{equation}
where the matrix $M$ is defined as
\begin{equation}
\label{eq:M-matrix}
    M_{ij} \; := \; E_{ij} + f^{k}\,_{ij} \, v_{k}~.
\end{equation}
Substituting these expressions back into the action yields the dual sigma model,
\begin{equation}
    \widehat{S} \= \int d\sigma^2 \left[ Q_{\mu \nu} \, \partial_{+} x^{\mu} \partial_{-} x^{\nu} + \left( \partial_{+} v_{i} + Q_{\mu i} \, \partial_{+} x^{\mu} \right) (M^{-1})_{ij} \left( \partial_{-} v_{j} - Q_{j\nu} \, \partial_{-} x^{\nu} \right) \right]~,
\end{equation}
from which the dual NS sector fields can be directly identified. Conversely, integrating out the Lagrange multipliers rather than the gauge fields enforces the flatness condition and recovers the original non-linear sigma model. In particular, the generalised Buscher rules take the form,
\begin{align}
\widehat{Q}_{\mu\nu} = Q_{\mu\nu} - Q_{\mu i} \,(M^{-1})_{ij} \, Q_{j\nu}\,, \qquad
\widehat{E}_{ij} = (M^{-1})_{ij}\,,\nonumber\\
\widehat{Q}_{\mu i} = Q_{\mu j} \, (M^{-1})_{ji}\,, \qquad
\widehat{Q}_{i\mu} = - (M^{-1})_{ij} \, Q_{j\mu}\,. 
\end{align}
Aditionally, at the quantum level, the dualisation induces a shift in the dilaton \cite{delaOssa:1992vci}. This arises from the functional determinant generated when the auxiliary gauge fields are integrated out. It is given by,
\begin{equation}
    \widehat{\Phi}(x,v) \= \Phi(x) - \frac{1}{2} \log(\, \det M)~
\end{equation}
which ensures conformal invariance of the dual sigma model at one loop.

The dual coordinates $v_i$ parameterise the Lie algebra $\mathfrak{g}$ of the isometry group and therefore define a non-compact vector space. 
For $\mathrm{SU(2)}$, one may introduce spherical coordinates,
\begin{equation}
\label{eq:lagmultipliersSU2}
v_1 = r_2 \sin\chi \cos\xi\,, \qquad v_2 = r_2 \sin\chi \sin\xi\,, \qquad v_3 = r_2 \cos\chi\,,
\end{equation}
while for $\mathrm{SL(2,\mathbb{R})}$ a hyperbolic parametrisation is appropriate,
\begin{equation}
\label{eq:lagmultipliersSL}
\widetilde{v}_1 = r_1 \sinh\widetilde{\xi} \,, \quad\quad
\widetilde{v}_2 = r_1 \cos\widetilde{\chi} \cosh\widetilde{\xi}\,, \quad  \quad \widetilde{v}_3 = r_1 \sin\widetilde{\chi} \cosh\widetilde{\xi}\,.
\end{equation}
In both cases, the dualisation replaces the original group manifold by a geometry involving a radial direction and a two-dimensional space, reflecting the non-compact nature of the dual geometry.

The transformation of the RR sector was performed in \cite{Sfetsos_2011} and follows from the fact that NATD acts differently on left- and right-moving worldsheet modes. As a consequence, the dual geometry admits two natural sets of frame fields, $\hat{e}_+$ and $\hat{e}_-$, which are related by a local Lorentz transformation, $\Lambda$.
While these frames describe the same target space metric, they differ by a Lorentz transformation, 
$\hat{e}_+ = \Lambda \hat{e}_-\,$,
that acts on spacetime spinors. In turn, this Lorentz transformation can be represented on spinorial quantities by a matrix $\Omega$, defined through its action on the gamma matrices,
\begin{equation}
\Omega^{-1} \Gamma^a \Omega = \Lambda^a{}_b \Gamma^b\,.
\end{equation}
Here, $\Gamma^{a}$ denote the ten-dimensional gamma matrices satisfying the Clifford algebra $\{ \Gamma^{a}, \Gamma^{b} \} = 2 \eta^{ab}$.
For dualisation along three directions, the solution to this condition can be written as\footnote{We thank Giel Boermans for pointing out the necessity of the factor
$\det(\eta_G)$ in the denominator, with $\det(\eta_G)=+1$ for $SU(2)$
and $\det(\eta_G)=-1$ for $SL(2,\mathbb{R})$.}
\begin{equation}\label{eq.NATD-lorentz-spinor}
    \Omega = \Gamma_{11} \frac{-\Gamma_{123}+ \zeta^a \Gamma_a}{\sqrt{1+\det(\eta_G)\zeta^2}},
\end{equation}
where $\eta_{G}$ denotes the metric on the dualised group directions, while $\zeta^a$ are auxiliary quantities determined by the dual coordinates and the matrix \eqref{eq:M-matrix}, which encode the corresponding local Lorentz transformation, and therefore depend on the original NS sector data. In this way, $\Omega$ provides a spinorial representation of the Lorentz transformation induced by the non-Abelian dualisation \cite{Sfetsos_2011,Itsios:2013wd}.

In the democratic formulation of Type II supergravity , the RR sector can be conveniently encoded in a bispinor,
\begin{equation}
P = \frac{e^{\Phi}}{2} \sum_n \frac{1}{n!} F_{a_1 \dots a_n} \Gamma^{a_1 \dots a_n}\,,
\end{equation}
where the sum runs over even (odd) values of $n$ in Type IIA (IIB), and $\Gamma^{i_1 \dots i_n}$ denotes the antisymmetrised product of gamma matrices. Under NATD, the transformation between the left- and right-moving frames induces a Lorentz transformation on spinors, implemented by the matrix $\Omega$. The dual RR bispinor is then given by
\begin{equation}
\widehat{P} = P \cdot \Omega^{-1}\,,
\end{equation}
where $\widehat{P}$ is defined analogously to $P$ in terms of the dual dilaton and RR fluxes. The same rule holds for Type IIA and Type IIB backgrounds. Since $\Omega$ has odd Clifford parity, $\widehat{P}$ and $P$ have opposite parity, so a dualisation along three directions maps Type IIA into Type IIB and vice versa.

\vspace{10pt}
In the following subsections we construct explicitly the dual backgrounds obtained by performing NATD along the sequences of isometries described above. As illustrated in Figure~\ref{Fig::Diagram}, the dualisation can follow two different routes, depending on whether the $\mathrm{SU}(2)$ or the $\mathrm{SL}(2,\mathbb{R})$ isometry is dualised first. This allows us to compare the compact and non-compact dualisations in the same set-up.

\subsection{Type A: the SL(2,$\mathbb{R}$) branch}
\label{sec:one}

The geometries obtained through NATD with respect to a freely acting 
$\mathrm{SL}(2,\mathbb{R})$ isometry on the AdS$_3$ factor of the solutions in \cite{Lozano:2019emq} 
were first constructed in \cite{Lozano:2021rmk}. 
The complete background reads,
\begin{equation}
\begin{split}
\text{d}{s}^2& \=\frac{u\sqrt{h_4 \, h_8}}{\Delta_1} \, r_1^2 \, \text{d}s_{\text{\,AdS}_2}^2+\sqrt{\frac{h_4}{h_8}} \, \text{d}s^2_{\text{\,CY}_2}+\frac{u\sqrt{h_4 \, h_8} }{\Delta} \, \text{d}s^2_{\text{\,S}^2}+\frac{\sqrt{h_4 \, h_8} }{u} \, (\text{d}\rho ^2 + \text{d}r_1^2)~,\\
e^{-2 \Phi}& \=\frac{\Delta\;\Delta_1}{2^8 \, h_4^2}~,\qquad
B_2 \=-\left(\frac{2 r_1^3 \, h_4 \, h_8}{\Delta_1}-\pi n\right)\text{vol}_{\text{ AdS}_2}+\frac{1}{2}\left(\rho-2\pi k-\frac{uu'}{\Delta}\right) \text{vol}_{\text{ S}^2} ~,\\
&\qquad\qquad \Delta \= 4 \, h_4 \, h_8 + (u')^2,  \qquad \Delta_1 \= 4 r_1^2 \, h_4 \, h_8-u^2~,\label{eq:typeINSNS}
\end{split}
\end{equation}
$n$ and $k$ are constants that we will clarify later. As in the parent solution, the functions $h_4$, $h_8$ and $u$ depend only on $\rho$, and the CY$_2$ enters the background only through its Ricci-flat metric and its volume form. Additionally, the background is supported by the RR fluxes,
\begin{equation}
\begin{split}
F_1 \=&-\frac{r_1 \, h'_8}{4} \, \text{d}r_1+\frac{1}{2^4}\left(4h_8+ \left(\frac{u u'}{h_4}\right)'\right) \, \text{d}\rho~,\\
F_3 \=&\frac{r_1^2 \, h_8}{8 \Delta_1}\left(\frac{h_8'\, u^2}{h_8} \, \text{d}r_1-r_1 \, h_4\left(4h_8+\left(\frac{u u'}{h_4}\right)'\right)\text{d}\rho\right)\wedge \text{vol}_{\text{ AdS}_2}\\
&+\frac{h_8}{8 \Delta}\left(\frac{h_4' \, u^2}{h_4} \, \text{d}\rho+r_1h_8\left(4h_4+\partial_\rho \left(\frac{u u'}{h_8}\right)'\right)\text{d}r_1\right)\wedge \text{vol}_{\text{ S}^2}~,\\
F_5 \=&\frac{1}{2^4}\left(4r_1 \, h_4' \, \text{d}r_1-\left(4h_4+\left(\frac{u u'}{h_8}\right)'\right)\text{d}\rho\right)\wedge \text{vol}_{\text{ CY}_2}\\
&-\frac{r_1^2 \, u^2 \, h_8^2}{2^4\Delta\;\Delta_1 }\left(4 r_1 \, h_4' \, \text{d$\rho $}+\left(4h_4+\left(\frac{u u'}{h_8}\right)'\right) \, \text{d$r_1$}\right)\wedge\text{vol}_{\text{ AdS}_2}\wedge \text{vol}_{\text{ S}^2}~,\\
%F_7=&-\frac{h_4}{8 \Delta}\left(\frac{h_8'u^2}{h_8}\text{d}\rho+r_1h_4\left(4h_8+\partial_\rho \left[\frac{u u'}{h_4}\right]\right)\text{d}r_1\right)\wedge \text{vol}_{\text{S}^2}\wedge \text{vol}_{\text{CY}_2}\\
%&-\frac{r_1^2h_4}{8 \Delta_1}\left(\frac{h_4'u^2}{h_4}\text{d}r_1-r_1h_8\left(4h_4+\partial_\rho \left[\frac{u u'}{h_8}\right]\right)\text{d}\rho\right)\wedge \text{vol}_{\text{AdS}_2}\wedge \text{vol}_{\text{CY}_2},\\
%F_9=&\frac{r_1^2u^2h_4^2}{4\Delta\;\Delta_1}\left(r_1h_8'\text{d}\rho+\frac{1}{4}\left(4h_8+ \partial_\rho \left[\frac{u u'}{h_4}\right]\right)\text{d}r_1\right)\wedge \text{vol}_{\text{AdS}_2}\wedge \text{vol}_{\text{CY}_2}\wedge \text{vol}_{\text{S}^2}.
\end{split}\label{eq:typeIRR}
\end{equation}
with the higher fluxes related to them by $F_7=-\star_{10}\,F_3,~F_9=\star_{10} \, F_1$. The metric and the dilaton are well defined for $\Delta_1=4 \, r_1^2 \, h_4 \, h_8-u^2 > 0$, that is, for
\begin{equation}
r_1>r_1^{(*)}=\frac{u}{2\sqrt{h_4 \, h_8}}~.\label{eq:r0natd}
\end{equation}
The radial coordinate $r_1$ is restricted to $r_2 \geq r_1^{(*)}$ , where $r_1^{(*)}$ denotes the lower endpoint of the $r_1$ coordinate range. The lower bound $r_1^{(*)}$ depends on the coordinate $\rho$, as it is determined by the background functions. This dependence will also play a role in the structure of the gauge patches introduced below. Near the lower bound $\Delta_1$ vanishes linearly in $r_1-r_1^{(*)}$. The expansion around this point is the local form of the fundamental string solution, with AdS$_2$ along the worldvolume and a harmonic function proportional to $\Delta_1$, which vanishes at $r_1=r_1^{(*)}$ instead of diverging as it would for a source of positive tension. We therefore interpret this end of the space as a smeared OF1-plane. The analogous lower endpoint in \cite{Ramirez:2021tkd} has the same local behaviour.

At this stage, it is important to clarify the role of the constants $n$ and $k$ appearing in the $B_2$ field in \eqref{eq:typeIRR}. These parameters arise from large gauge transformations of the NSNS two-form. In general, the $B_2$ field is defined up to large gauge transformations of the form
\begin{equation}
\label{eq:lgtNATD}
B_2 \;\rightarrow\; B_2 + \Delta B_2 \,,
\end{equation}
where $\Delta B_2$ is a closed two-form implementing a large gauge transformation. 
In the present setup, the large gauge transformations of the $\mathrm{S}^2$  component of $B_2$ are quantized due to the compactness of $\mathrm{S}^2$, while the corresponding transformation along $\mathrm{AdS}_2$ is fixed by consistency with the definition of charges, in line with the results of \cite{Ramirez:2021tkd, Lozano:2020sae}.

In the present background, the NSNS two-form contains two independent components, one supported on $\mathrm{AdS}_2$ and the other on S$^2$, namely
\begin{equation}
B_2 = B_2^{e}\,\text{vol}_{\mathrm{AdS}_2} + B_2^{m}\,\text{vol}_{\mathrm{S}^2} \,.
\end{equation}
Accordingly, the corresponding large gauge transformations can be implemented independently on the two-dimensional subspaces. With the conventions adopted here, they take the form
\begin{equation}
    \Delta B_2  =\pi n\,\text{vol}_{\mathrm{AdS}_2}-\pi k\,\text{vol}_{S^2},
    \qquad n,k\in\mathbb{Z}.
\label{eq:LGTTypeI}
\end{equation}
The integers $n$ and $k$ therefore label the electric and magnetic gauge patches, respectively.

To determine the corresponding fundamental regions, we impose the usual condition separately on the two components of $B_2$,
\begin{equation}
    b_{\rm e}\equiv \frac{1}{4\pi^2} \left| \int_{\mathrm{AdS}_2} B_2\right|\in[0,1), \quad\quad\quad b_{\rm m}\equiv \frac{1}{4\pi^2}\left|\int_{S^2} B_2 \right| \in[0,1).
\label{eqref:boundB2}
\end{equation}
For the non-compact $\mathrm{AdS}_2$ factor we use the regularised volume
$\text{vol}_{\mathrm{AdS}_2}=4\pi$, following the prescription of ~\cite{Ramirez:2021tkd,Lozano:2020sae}. With this choice, both transformations in
\eqref{eq:LGTTypeI} shift the corresponding quantities $b_{\rm e}$ and $b_{\rm m}$ by one unit.

We begin with the magnetic component of $B_2$, which determines the gauge patches along the $\rho$ direction. The corresponding endpoints $\rho^{(k)}$ are fixed implicitly by
\begin{equation}
    \frac{1}{2}
    \left(
        \rho^{(k)}
        -
        \frac{
            u(\rho^{(k)})\,u'(\rho^{(k)})
        }{
            \Delta(\rho^{(k)})
        }
    \right)
    =
    \pi k .
\label{eq:rhoPatchesTypeI}
\end{equation}
Once a magnetic patch labelled by $k$ has been selected, the functions
$h_4$, $h_8$, $u$, and consequently $\Delta$, are fixed throughout the interval
$\rho\in[\rho^{(k)},\rho^{(k+1)}]$.

Within each such magnetic patch, the electric component of $B_2$ determines a second family of gauge-patch boundaries along the $r_1$ direction. These are defined by
\begin{equation}
    \frac{
        2\bigl(r_1^{(n)}\bigr)^3
        h_4(\rho)\,h_8(\rho)
    }{
        \Delta_1\bigl(r_1^{(n)},\rho\bigr)
    }
    =
    \pi n .
\label{eq:r1PatchesTypeI}
\end{equation}
Thus, for each magnetic interval labelled by $k$, the electric large gauge transformation introduces a family of boundaries $r_1^{(n)}(\rho)$. The pair of integers $(k,n)$ therefore labels the gauge patches associated with the magnetic and electric components of $B_2$, respectively. Since $r_1^{(n)}$ depends in general on $\rho$, these patches need not form a rectangular lattice in the $(\rho,r_1)$ plane.

For field theory purposes, it is convenient to introduce the Page fluxes, $\widehat F_p = (e^{-B_2} \wedge F)_p$, which are sensitive to the choice of $B_2$ and provide the appropriate quantised charges characterising the background, 
\begin{align}
\label{eq:PagestypeI}
\widehat{F}_1  \=& \widehat{F}_{1,\rho}-\frac{1}{4}\, r_1\, h_8' \, dr_1\, 
       ~,
\nonumber\\
\widehat{F}_3  \=&-
       n \pi\, \widehat{F}_{1,\rho}\wedge \mathrm{vol}_{\textrm{ AdS}_2}-\frac{1}{8}\, r_1 \, h_8'
       (r_1 - 2 n \pi)\, dr_1\wedge \mathrm{vol}_{\textrm{ AdS}_2}\, 
         \nonumber\\
        &-\widehat{F}_{3,\rho}\wedge \mathrm{vol}_{\textrm{S}^2}- \frac{1}{8}
        \, r_1\,  \left( h_8 - (\rho - 2 k \pi)\, h_8' \right)dr_1
       \wedge \mathrm{vol}_{\, \mathrm{S}^2}~,
\nonumber\\
\widehat{F}_5  \=&  n \pi\, \,  \widehat{F}_{3,\rho}\wedge \mathrm{vol}_{\, \mathrm{S}^2}\wedge \mathrm{vol}_{\, \mathrm{AdS}_2}-\frac{1}{16}
       \, r_1 (r_1 -2 n \pi)\,  \left( h_8 - (\rho - 2 k \pi)\, h_8' \right)dr_1\wedge \mathrm{vol}_{\, \mathrm{AdS}_2} \wedge \mathrm{vol}_{\, \mathrm{S}^2} \nonumber\\   
        &+ \widehat{F}_{5,\rho}\wedge \mathrm{vol}_{ \, \mathrm{CY}_2}- 
        \, \frac{1}{4} r_1\, h_4'\, dr_1\wedge \mathrm{vol}_{ \, \mathrm{CY}_2}~,
\nonumber\\
\widehat{F}_7  \=& 
        - n \pi\, \widehat{F}_{5,\rho}\wedge \mathrm{vol}_{ \, \mathrm{CY}_2}\wedge \mathrm{vol}_{ \, \mathrm{AdS}_2}-\frac{1}{8}\, r_1 \, h_4'(r_1 - 2 n \pi)\, dr_1\wedge \mathrm{vol}_{ \, \mathrm{AdS}_2} \wedge \mathrm{vol}_{\, \mathrm{CY}_2}\nonumber\\
    &-\widehat{F}_{7,\rho}\wedge \mathrm{vol}_{ \, \mathrm{S}^2} \wedge \mathrm{vol}_{\, \mathrm{CY}_2}- \frac{1}{8}r_1  \left( h_4 - (\rho - 2 k \pi)\, h_4' \right)dr_1\wedge \mathrm{vol}_{\, \mathrm{S}^2} \wedge \mathrm{vol}_{ \, \mathrm{CY}_2} ~,
\nonumber\\
\widehat{F}_9  \=&n\pi\,\widehat{F}_{7,\rho}\wedge \mathrm{vol}_{\, \mathrm{CY}_2}
      \wedge \mathrm{vol}_{ \, \mathrm{AdS}_2}\nonumber\\
      &-\frac{1}{16}
        r_1 (r_1 - 2n \pi)\,  \left( h_4 - (\rho - 2 k \pi)\, h_4' \right)dr_1\wedge \mathrm{vol}_{ \, \mathrm{AdS}_2} \wedge \mathrm{vol}_{ \, \mathrm{S}^2} \wedge \mathrm{vol}_{\, \mathrm{CY}_2}~.
\end{align}
with
\begin{align}
    \widehat{F}_{1,\rho} & \=\frac{1}{16}\, d\rho \left( 4 h_8
      + \!\left(\frac{u\,u'}{h_4}\right)' \right),\quad\quad \widehat{F}_{5,\rho}  \= \frac{1}{16}\,  \left( 4 h_4
        + \left(\frac{u\,u'}{h_8}\right)' \right)d\rho,\nonumber\\
      \widehat{F}_{3,\rho} & \=\frac{1}{32}\,  \left(
           (\rho - 2 k \pi)\left( 4 h_8
           + \left(\frac{u\,u'}{h_4}\right)'\right)
           - u\, \!\left(\frac{u}{h_4}\right)'
        \right)d\rho,\nonumber\\
      \widehat{F}_{7,\rho} & \= \frac{1}{32}\,  \left(
           (\rho - 2 k \pi)\left( 4 h_4
           + \left(\frac{u\,u'}{h_8}\right)'\right)
           - u\, \!\left(\frac{u}{h_8}\right)'
        \right)d\rho 
\end{align}

The Page forms provide the quantised fluxes from which the brane charges of the solution can be extracted. In our conventions, the magnetic charge carried by a Dp-brane is defined as
\begin{equation}
Q^{m}_{\mathrm{D}p}=\frac{1}{(2\pi)^{7-p}}\int_{\Sigma_{8-p}}\widehat{F}_{8-p},\label{eq:magneticPage}
\end{equation}
where $\Sigma_{8-p}$ is an $(8-p)$-dimensional cycle transverse to the corresponding brane. It will also be useful to keep track of the electric components of the Page fluxes. We do so through the quantities
\begin{equation}
Q^{e}_{\mathrm{D}p}=\frac{1}{(2\pi)^{p+1}}\int_{\mathrm{AdS}_{2}\times\Sigma_{p}}\widehat{F}_{p+2},
\label{eq:electricPage}
\end{equation}
where $\Sigma_p$ denotes a $p$-cycle wrapped by the Dp-brane and, as in the discussion of the large gauge transformations above, the integral over $\mathrm{AdS}_{2}$ is understood in the regularised sense, with $\textrm{vol}_{\mathrm{AdS}_{2}}=4\pi$.

The charge of the NS5-branes is obtained in the same way from $H_3=dB_2$,
\begin{equation}
Q_{\mathrm{NS5}}=\frac{1}{4\pi^2}\int_{\Sigma_3}H_3~,
\label{eq:NSchargesTypeI}
\end{equation}
with $\Sigma_3=[\rho^{(k)},\rho^{(k+1)}]\times\mathrm{S}^2$. Using \eqref{eq:rhoPatchesTypeI} one finds $|Q_{\mathrm{NS5}}|=1$ in each patch, so a new NS5-brane is created each time we cross $\rho=\rho^{(k)}$. Similarly, the regularised integral of $H_3/4\pi^2$ over $\mathrm{AdS}_2\times[r_1^{(n)},r_1^{(n+1)}]$ at fixed $\rho$ equals one in each patch by \eqref{eq:r1PatchesTypeI}. In \cite{Ramirez:2021tkd,Lozano:2020sae} this was interpreted as the creation of a fundamental string each time $r_1=r_1^{(n)}$ is crossed.

The structure of the Page forms in \eqref{eq:PagestypeI} naturally distinguishes two families of magnetic charges. The components containing $dr_{1}$ define the charges associated with D1-, D3-, D5- and D7-branes whose transverse cycles contain the $r_{1}$ direction. Their charges are
\begin{align}
\label{eq:DpmagneticchargestypeI}
    &Q_{1}=\frac{h_4-(\rho-2\pi k)h_4'}{8\pi}\int r_1 dr_1,\qquad  Q_{5}=\frac{h_8-(\rho-2\pi k)h_8'}{8\pi}\int r_1 dr_1,\nonumber\\
     &Q_{3}=\frac{1}{4}\int \int r_1 h_4'' \;d\rho\; dr_1,\qquad\qquad\quad  Q_{7}=\frac{1}{4}\int \int r_1 h_8'' \;d\rho\; dr_1 .
\end{align}
At this stage, the integration ranges entering these expressions remain unspecified. A global completion of the geometry, introduced in Section~\ref{sec:FTtypeI}, will fix the corresponding integration domains and allow the charges to be evaluated explicitly.

A complementary set of magnetic charges is obtained from the Page fluxes supported along the $\rho$ direction. We denote the corresponding objects by Dp$'$, where the prime distinguishes their orientation from that of the Dp-branes associated with the cycles containing $r_{1}$. Their magnetic charges are
\begin{align}
\label{eq:DpprimemagneticchargestypeI}
    &Q^m_{D1'}=-\frac{1}{\pi}\int \widehat{F}_{7,\rho},\qquad  Q^m_{D5'}=-\frac{1}{\pi}\int \widehat{F}_{3,\rho},\nonumber\\
     &Q^m_{D3'}=\int \widehat{F}_{5,\rho},\qquad  Q^m_{D7'}=\int \widehat{F}_{1,\rho}.
\end{align}
The distinction between the primed and unprimed branes is therefore geometrical: the two families are obtained by integrating different components of the same Page forms over cycles with different orientations in the $(\rho,r_{1})$ plane. As we will see in later sections, these sets of branes will play different roles.

We now evaluate the quantities \eqref{eq:electricPage}. Using the corresponding components of the fluxes in \eqref{eq:PagestypeI}, we obtain
\begin{align}
\label{eq:DpelectricchargestypeI}
   &Q^e_{1}=-n\int \widehat{F}_{1,\rho},\qquad  Q^e_{5}=-n\int \widehat{F}_{5,\rho},\nonumber\\
     &Q^e_{3}=\frac{n}{\pi}\int \widehat{F}_{3,\rho},\qquad  Q^e_{7}=\frac{n}{\pi}\int \widehat{F}_{7,\rho},
\end{align}
The explicit dependence on $n$ originates from the electric large gauge transformations of $B_2$. Comparing \eqref{eq:DpelectricchargestypeI} with the magnetic charges of the primed branes in \eqref{eq:DpprimemagneticchargestypeI}, we find
\begin{align}
\label{eq:DpelectricmagneticchargestypeI}
    &Q^e_{D1}=-nQ^m_{D7'},\qquad\qquad\qquad Q^e_{D3}=-nQ^m_{D5'},\nonumber\\ &Q^e_{D5}=-nQ^m_{D3'},\qquad\qquad\qquad Q^e_{D7}=-nQ^m_{D1'}.
\end{align}
Thus, the electric charge of each unprimed Dp-brane is related to the magnetic charge of the complementary D$_{(8-p)'}$-brane. The primed charges are not all independent, however. As we show in Section~\ref{sec:FT}, the D5$'$ and D1$'$ charges are induced by $B_2$ on the D7$'$ and D3$'$ branes, so only the relations for the D1 and D5 branes carry independent information. Their physical interpretation will be discussed in Section~\ref{sec:dualQM}.

\subsubsection{Holographic central charge}
\label{subsub:cchol}

For backgrounds containing an $\mathrm{AdS}_2$ factor, the notion of central charge is more subtle than in higher-dimensional AdS/CFT. Since the dual theory is a conformal quantum mechanics, the quantity extracted holographically should be interpreted as a measure of the ground-state degeneracy rather than as a conventional Virasoro central charge \cite{Lozano:2020txg}. Nevertheless, following the standard prescription in the literature \cite{Macpherson:2014eza, Bea:2015fja}, one can define an effective two-dimensional Newton constant by reducing the ten-dimensional action on the internal space.
The relevant quantity is the warped internal volume,
\begin{equation}
V_{\text{int}} \;=\; \int d^{8}x \, \sqrt{e^{-4\Phi}g_8}\,  ~,
\end{equation}
where $g_8$ denotes the metric of the internal space (i.e. the part transverse to $\mathrm{AdS}_2$). Upon dimensional reduction, this determines the effective two-dimensional Newton constant as
\begin{equation}
\frac{1}{G_{N,2}} \;=\; \frac{V_{\text{int}}}{G_{N,10}}~.\nonumber
\end{equation}

In analogy with higher-dimensional cases, we define the holographic central charge of the dual conformal quantum mechanics as
\begin{equation}
\label{eq:ccholAdS2}
c_{\text{hol}} \;=\; \frac{3}{4\pi \, G_{N,2}}
\;=\; \frac{3\,V_{\text{int}}}{4\pi \, G_{N,10}} ~.
\end{equation}
Up to possible rescalings depending on conventions, this quantity is supposed to capture the effective number of vacuum states of the dual SCQM.
For the Type A background, the central charge reads
\begin{align}
    c_{\text{hol}} \;=\;\frac{3}{2^7\pi^2}\int{\Delta_1}\;d\rho \;dr_1 ~.
\end{align}
The integrand is positive in the region \eqref{eq:r0natd} where the solution is well defined. As for the charges, the integration ranges are only fixed once the geometry is globally completed, and we come back to this expression in Section~\ref{sec:dualQM}, where it is compared with the central charge of the dual quantum mechanics.

\subsection{The SU(2) branch and the common double-dual background}
\label{sec:twoandthree}

In this subsection we first perform SU$(2)$ NATD on the parent geometry, which gives the Type B background. We then dualise the SL$(2,\mathbb{R})$ isometry of its AdS$_3$ factor and obtain the Type C geometry, which completes the diagram in Figure~\ref{Fig::Diagram}.

\subsubsection{Type B: SU(2) NAT-dual geometry}
\label{sec:two}
For this branch we take $\mathrm{M}_4=\mathrm{S}^3\times\mathbb{R}$ in the parent geometry and dualise with respect to the SU$(2)$ isometry acting freely on the S$^3$. Unlike in the Type A case, where NATD acts on the non-compact AdS$_3$ factor, here it acts on a compact internal space and the AdS$_3$ is left untouched. The result is a Type IIB background that preserves the same amount of supersymmetry as the parent solution. Indeed, the SU$(2)_R$ R-symmetry is realised on the S$^2$ and is not affected by the dualisation, which ensures that the supersymmetry associated to this R-symmetry is preserved under NATD \cite{Kelekci:2014ima}.

In the spherical coordinates introduced in Section~\ref{sec:NATD}, the S$^3$ is replaced by the radial direction $r_2$ and a two-sphere $\widetilde{\mathrm{S}}^2$ parametrised by $(\chi,\xi)$. The NSNS sector reads
\begin{equation}
\label{TypeIIBckgn}
    \begin{split}
ds^2 & = 
\frac{u}{\sqrt{h_4 h_8}} ds_{\, \mathrm{AdS}_3}^2 + \sqrt{h_4 h_8} \, \left(  \frac{u}{\Delta} ds_{ \mathrm{S}^2}^2 + \frac{4  r^2 \, r_2^2}{\Delta_2}  ds_{\, \widetilde{\mathrm{S}}^2}^2 \right)+ \frac{\sqrt{h_4 h_8}}{u} d\rho^2 
+  \sqrt{ \frac{h_4}{h_8} }dr^2 
+ \frac{4}{r^2}   \sqrt{ \frac{h_8}{h_4} }dr_2^2,\\
e^{-2 \Phi} &= \frac{r^2 \Delta \, \Delta_2}{2^8 \, u}, \qquad B_2 = \frac{1}{2} \left( \frac{u \, u'}{\Delta} - \rho \right)  \mathrm{vol}_{\, \mathrm{S}^2} -\frac{2^4 \, r_2^3 \, h_8}{\Delta_2}  \mathrm{vol}_{\, \widetilde{\mathrm{S}}^2}.
  \end{split}
\end{equation}
where $ \Delta_2 = r^4 \, h_4 + 16 \, r_2^2 \, h_8$ controls the geometry of the dual two-sphere $\widetilde{\mathrm{S}}^2$, the dilaton and $B_2$ field.  
As in the parent geometry, the functions $h_4$, $h_8$ and $u$ depend only on the coordinate $\rho$ and continue to satisfy the conditions discussed in Appendix \ref{app:B-Appendix}. 

The background is supported by the following RR fluxes, 
\begin{equation}
\begin{split}
    F_1 \=& \frac{1}{8} r^3\, h_4' \, dr  + r_2 \, \, h_8' \, dr_2 ~,\\
    F_3 \=&- \left( \, \frac{r^3 (h_4 \, \Delta - u \, h_4' \, u')}{16 \, \Delta} \, dr + \frac{r_2 (h_8 \, \Delta - u \, h_8' \, u')}{2 \, \Delta} \, dr_2 \, \right) \wedge \mathrm{vol}_{\, \mathrm{S}^2}\\
&+ \left( \, \frac{2\, r^3 \, r_2^3 \, h_8 \, h_4'}{\Delta_2} \, dr
- \frac{r^4 \, r_2^2 \, h_4 \, h_8'}{\Delta_2} \, dr_2 \, \right) \wedge \mathrm{vol}_{\, \widetilde{\mathrm{S}}^2}~,\\
F_5 \=&
\Biggl( \,
 \frac{r_2}{2 \, h_4^2} \bigl( h_4 \, \Delta - u \, h_4' \, u' \bigr) \, dr_2 - \frac{ r^3 \, \bigl( h_8 \, \Delta - u \, h_8' \, u' \bigr)}{16 \, h_8^2} \, dr \,
\Biggr) 
\wedge d\rho \wedge \mathrm{vol}_{\,\mathrm{AdS}_3} \\
&+  \Biggr( \,
\frac{ r^4 \, r_2^2 \, h_4 \bigl( h_8 \, \Delta - u \, h_8' \, u' \bigr)}{2 \, \Delta_2 \, \Delta} \, dr_2 
- \frac{ r^3 \, r_2^3 \, h_8 \bigl( h_4 \, \Delta - u \, h_4' \, u' \bigr)}{ \Delta_2 \, \Delta} \, dr \,
\Biggr) 
\wedge \mathrm{vol}_{\, \mathrm{S}^2} \wedge \mathrm{vol}_{\, \widetilde{\mathrm{S}}^2}.
\end{split}
\end{equation}
This is the Type B branch of Figure~\ref{Fig::Diagram}. Its $\mathrm{AdS}_3$ factor provides the $\mathrm{SL}(2,\mathbb{R})$ isometry used in the second dualisation below.

\noindent\textbf{Holographic central charge.}
 For the AdS$_3$ solutions of \eqref{TypeIIBckgn} we follow the prescription of \cite{Macpherson:2014eza,Bea:2015fja}, by which the holographic central charge is given by
\begin{equation}
c_{\text{hol}} \= \frac{3\,V_{\text{int}}}{2\,G_{N,10}}~, %\qquad V_{\text{int}} \= \int d^7\theta\, \sqrt{e^{-4\Phi}\,\det g_{\text{int}}\;\frac{u}{\sqrt{h_4h_8}}}~,
\end{equation}
with $V_{\text{int}}$ denoting the volume of the seven-dimensional internal space. For the background \eqref{TypeIIBckgn}, this gives
\begin{equation}
V_{\text{int}} \= \frac{\pi^2}{2}\int d\rho\, dr\, dr_2\; r^3 r_2^2\, h_4 h_8~,
\qquad
c_{\text{hol}} \= \frac{3}{2^5\pi^4}\int d\rho\, dr\, dr_2\; r^3 r_2^2\, h_4 h_8~,
\end{equation}
where we have set $G_{N,10}=8\pi^6$. Since $\mathrm{M}_4=\mathbb{R}^4$ is non-compact, the integral over $r$ is not finite. The range of the dual coordinate $r_2$ is not fixed by the local NATD transformation, so the expression above should be understood as a local expression in $r_2$, as was the case for $r_1$ in the Type A.

\subsubsection{Type C: SL(2,$\mathbb{R}$)+ SU(2) NAT-dual geometry}
\label{sec:three}
The Type C background can be obtained either by SL$(2,\mathbb{R})$ dualisation of the Type B geometry or by SU$(2)$ dualisation of the Type A geometry of Section~\ref{sec:one} with $\mathrm{M}_4=\mathbb{R}^4$, and we have checked explicitly that both routes give the same result. The SL$(2,\mathbb{R})$ dual coordinates build the $(r_1,\mathrm{AdS}_2)$ sector, controlled by $\Delta_1$ as in Type A, while the SU$(2)$ ones build the $(r_2,\widetilde{\mathrm{S}}^2)$ sector, controlled by $\Delta_2$ as in Type B. The NSNS sector reads
\begin{align*}
ds^2 & \=  u\sqrt{h_4 \, h_8} \left( 
\frac{r_1^2 }{\Delta_1}ds^2_{\text{ AdS}_2}+ \frac{1}{\Delta}ds^2_{\text{ S}^2}
\right) + \frac{4\, r^2 \, r_2^2 \sqrt{h_4 \, h_8}}{\Delta_2} ds^2_{ \, \widetilde{\, \text{S}}^2}+ \frac{\sqrt{h_4 \, h_8}}{u} \left( dr_1^2 + d\rho^2 \right)\\
&\quad +   \sqrt{ \frac{h_4}{h_8} }dr^2 
+ \frac{4}{r^2}   \sqrt{ \frac{h_8}{h_4} }dr_2^2~,\qquad\qquad e^{-2\Phi} \= \frac{r^2 \, \Delta_2 \, \Delta_1 \, \Delta}{2^{14} \left( h_4 \, h_8 \right)^{3/2}} ~,\\
B_2 & \=-\frac{2 \, r_1^3 \, h_4 \, h_8}{\Delta_1} \text{vol}_{\text{ AdS}_2}+\frac{1}{2}\left(-\rho+\frac{ u u'}{\Delta}\right)\text{vol}_{\text{ S}^2} - \frac{16 \, r_2^3 \, h_8}{\Delta_2} \text{vol}_{\widetilde{\, \text{S}}^2} .
\end{align*}
Notice that the NSNS two-form contains three independent contributions, respectively associated with the original $\mathrm{S}^2$ and with the dual spaces generated by each of the two non-Abelian transformations.

The background is supported by the following RR fluxes, with $F_0 = 0$, together with, 
\begin{align*}
F_2 & \= \frac{h_4'}{32} \, r^3 \, r_1 \, dr \wedge dr_1 + \frac{h_8'}{4} \, r_1 \, r_2 \, dr_2 \wedge dr_1  - \frac{r_2}{16} \left( 4 h_8 + u' \partial_{\rho} \left( \frac{u}{h_4} \right) \right) dr_2 \wedge d\rho \nonumber\\
&\quad - \frac{r^3}{128}  \left( 
4 h_4 + u'  \partial_{\rho} \left( \frac{u}{h_8} \right) \right) dr \wedge d\rho ~,\\
F_4 & \= -\frac{1}{64\,\Delta_1} \Bigg[r_1^2 \, u^2 \left( h_4' \, r^3 \, dr \wedge dr_1  + 8  h_8' \, r_2 \, dr_2 \wedge dr_1 \right) - h_4 \, h_8 \left( r^3 \, r_1^3 \left( 4 \, h_4 + \partial_\rho \left( \frac{u \, u'}{h_8} \right) \right) dr \wedge d\rho  \right.\nonumber\\
&\qquad \qquad \qquad \left.+ 8\, r_1^3 \, r_2 \left( 4\, h_8 + \partial_\rho \left( \frac{u \, u'}{h_4} \right) \right) dr_2 \wedge d\rho
\right) \Bigg] \wedge \text{vol}_{\, \mathrm{AdS}_2} \nonumber\\
&\quad - \frac{r^3 \, r_2^2 }{8 \, \Delta_2} \Bigg[ \, \frac{h_4}{2} r \left( 4\, h_8 + \partial_\rho \left( \frac{u\, u'}{h_4} \right) \right) dr_2 \wedge d\rho 
-2 \, r_1 \left( r \, h_4 \, h_8' \, dr_2 \wedge dr_1 - 2\, r_2 \, h_8 \, h_4' \, dr \wedge dr_1  \right) \nonumber\\
&\quad - r_2 \, h_8  \left( 4\, h_4 + \partial_\rho \left( \frac{u\, u'}{h_8} \right) \right) dr \wedge d\rho \,
\Bigg] \wedge \text{vol}_{ \, \widetilde{ \mathrm{ S}}^2} \nonumber\\
&\quad - \frac{1}{64 \, \Delta} \Bigg[
\frac{u^2}{h_4 \, h_8} \left( 
8\, h_8^2 \, h_4'\, r_2 \, dr_2 \wedge d\rho + r^3\, h_4^2 \, h_8' \, dr \wedge d\rho  \right)+ r^3 \, h_4^2 \, r_1 \left( 
4 h_8 + \partial_\rho \left( \frac{u u'}{h_4} \right) \right) dr \wedge dr_1  \nonumber\\
&\quad 
+ 8 \, h_8^2 \, r_1 \, r_2  \left( 
4 \, h_4 + \partial_\rho \left( \frac{u \, u'}{h_8} \right) \right) dr_2 \wedge dr_1 
\Bigg] \wedge \text{vol}_{\, \mathrm{S}^2} ~.
\end{align*}
This background can also be obtained in a single dualisation with respect to SL$(2,\mathbb{R})\times$SU$(2)$, see Appendix~\ref{sec:oneStep}. The Type C background preserves the same amount of supersymmetry as the Type A and Type B solutions, since both groups leave the Killing spinors invariant, as noted in \cite{Lozano:2021rmk,Ramirez:2021tkd} for SL$(2,\mathbb{R})$ and in Section~\ref{sec:two} for SU$(2)$.

Although the parent solution has a non-vanishing Romans mass, $F_0=h_8'$, the Type C background is a solution of massless Type IIA supergravity. The D8-branes of the parent become D7-branes in the Type A and Type B backgrounds and D6-branes in the Type C background, in analogy with two Abelian T-dualities along the worldvolume of a D8-brane.

%For AdS$_3\times$S$^3\times$T$^4$ supported by RR flux, the dualisation with respect to SL$(2,\mathbb{R})\times$SU$(2)$ gives the background obtained, after an analytic continuation, as a limit of the $\lambda$-deformed SL$(2,\mathbb{R})\times$SU$(2)$ model \cite{Sfetsos:2014cea,Itsios:2023kma}. The $\lambda$-deformation requires AdS$_3$ and S$^3$ to be unwarped group manifolds, while non-Abelian T-duality only uses their isometries and therefore also applies to backgrounds in which they are warped over other directions, such as the parent solution of Appendix~\ref{app:B-Appendix}.

\vspace{3pt}

\noindent\textbf{Holographic central charge.}
Since the Type C background contains an AdS$_2$ factor, we use the prescription of Section~\ref{subsub:cchol}, with the central charge given by \eqref{eq:ccholAdS2}. For the eight-dimensional internal space this gives
\begin{equation} \label{eq:VintChol}
V_{\text{int}} \= \frac{\pi^2}{2^7}\int d\rho\, dr\, dr_1\, dr_2\; r^3 r_2^2\, \Delta_1~,
\qquad
c_{\text{hol}} \= \frac{3}{2^{12}\pi^5}\int d\rho\, dr\, dr_1\, dr_2\; r^3 r_2^2\, \Delta_1~.
\end{equation}
 As for the Type B background, the integral over $r$ is not finite, and the ranges of the dual coordinates $r_1$ and $r_2$ are not fixed by the local NATD background, so \eqref{eq:VintChol} should be understood as local expressions in both coordinates.
\vspace{10pt}

The Type B and Type C backgrounds complete the two dualisation chains of Figure~\ref{Fig::Diagram}, and we have included their central charges for completeness. A detailed holographic analysis is most natural for the Type A solution, where $\mathrm{M}_4$ can be taken to be a compact CY$_2$, and this is the case we study in the following Sections.

\section{AdS$_2\times$S$^2$ geometries in Type IIB supergravity}
\label{sec:FT}
In this section we consider the supersymmetric Type IIB solutions with $\mathrm{AdS}_2\times\mathrm{S}^2\times\mathrm{CY}_2$ factors constructed with $G$-structure techniques in Section~5.2 of \cite{Legramandi:2023fjr}. We summarise their supergravity description in our conventions, together with the Page fluxes, Bianchi identities and charges that we shall need in Section~\ref{sec:dualQM} to propose the dual superconformal quantum mechanics.

We restrict to the sector in which the CY$_2$ enters only through its Ricci-flat metric and its volume form, and the background does not depend on its coordinates, so that it depends only on $(x_1,x_2)$, through the functions $h_3(x_1,x_2)$, $h_7(x_1,x_2)$ and $v(x_1,x_2)$. As we show in Section~\ref{sec:FTtypeI}, for a suitable choice of these functions this class contains the Type A backgrounds obtained via SL$(2,\mathbb{R})$ NATD. The NSNS sector reads 
\begin{align}
\label{eq:NSNSNiallsol}
ds^2 & \= 
v\,\sqrt{h_3 \, h_7}\left( \frac{1}{\Xi_1} \,
ds^2_{\textrm{ AdS}_2}
+ \frac{1}{\Xi_2} \, ds^2_{\textrm{ S}^2}\right)
+ \sqrt{\frac{h_3}{h_7}} \, ds^2_{\, \mathrm{ CY}_2}+\frac{\sqrt{h_3 \, h_7}}{v}
\left(dx_1^2 + dx_2^2\right)~,\nonumber\\
e^{-2\Phi}& \=\frac{c_0^2 \, \Xi_1 \,  \Xi_2}{h_3^{\,2}}~,\qquad\qquad \Xi_1 = h_3 \, h_7 - \dot v^{\,2}~, 
\qquad \qquad
\Xi_2 = h_3 \, h_7 + v'^{\,2}~,\nonumber\\
B_2 & \=
-\left((x_2-n\pi)+
\frac{v \, \dot v }{\Xi_1}
\right)\mathrm{vol}_{\, \mathrm{AdS}_2}+\left((x_1-k\pi)-
\frac{  v \, v'}{\Xi_2}
\right)\mathrm{ \,vol}_{\, \mathrm{ S}^2}~.
\end{align}
where primes denote derivatives with respect to $x_1$, and overdots derivatives with respect to $x_2$. As for the Type A background, the integers $k$ and $n$ label the large gauge transformations of the magnetic and electric components of $B_2$ respectively, see Section~\ref{sec:one}.
The charge analysis is most conveniently expressed in terms of the Page fluxes, $\widehat{F}=e^{-B_2}\wedge F$,
\begin{align}
\widehat{F}_1
&= c_0 \Bigg[
\dot{h}_7\,\mathrm{d}x_1-h_7'\,\mathrm{d}x_2
+\mathrm{d}\left(\frac{\dot{v}v'}{h_3}\right)
\Bigg],
\nonumber\\
\widehat{F}_3
&= c_0\,\mathrm{vol}_{\mathrm{AdS}_2}\wedge
\Bigg[
(x_2-n\pi)
\left(\dot{h}_7\,\mathrm{d}x_1-h_7'\,\mathrm{d}x_2\right)
-h_7\,\mathrm{d}x_1
+\mathrm{d}\left(
\frac{\big((x_2-n\pi)\dot{v}-v\big)v'}{h_3}
\right)
\Bigg]
\nonumber\\
&\quad
-c_0\,\mathrm{vol}_{\mathrm{S}^2}\wedge
\Bigg[
(x_1-k\pi)
\left(\dot{h}_7\,\mathrm{d}x_1-h_7'\,\mathrm{d}x_2\right)
+h_7\,\mathrm{d}x_2
+\mathrm{d}\left(
\frac{\big((x_1-k\pi)v'-v\big)\dot{v}}{h_3}
\right)
\Bigg],\nonumber\\
\widehat{F}_5
&= c_0\,\mathrm{vol}_{\mathrm{AdS}_2}
\wedge\mathrm{vol}_{\mathrm{S}^2}\wedge
\Bigg[
(x_1-k\pi)\big(h_7-(x_2-n\pi)\dot{h}_7\big)
\,\mathrm{d}x_1-(x_2-n\pi)\big(h_7-(x_1-k\pi)h_7'\big)
\,\mathrm{d}x_2\nonumber\\
&-\mathrm{d}\left(
\frac{\big(v-(x_2-n\pi)\dot{v}\big)
      \big(v-(x_1-k\pi)v'\big)}{h_3}
\right)
\Bigg]-c_0\,\mathrm{vol}_{\mathrm{CY}_2}\wedge
\Bigg[
\dot{h}_3\,\mathrm{d}x_1-h_3'\,\mathrm{d}x_2
+\mathrm{d}\left(\frac{\dot{v}v'}{h_7}\right)
\Bigg],
\nonumber
\end{align}
\begin{align}
\widehat{F}_7
&= c_0\,\mathrm{vol}_{\mathrm{AdS}_2}
\wedge\mathrm{vol}_{\mathrm{CY}_2}\wedge
\Bigg[
\big(h_3-(x_2-n\pi)\dot{h}_3\big)\,\mathrm{d}x_1
+(x_2-n\pi)h_3'\,\mathrm{d}x_2+\mathrm{d}\left(
\frac{\big(v-(x_2-n\pi)\dot{v}\big)v'}{h_7}
\right)
\Bigg]
\nonumber\\
&\quad
+c_0\,\mathrm{vol}_{\mathrm{S}^2}
\wedge\mathrm{vol}_{\mathrm{CY}_2}\wedge
\Bigg[
(x_1-k\pi)\dot{h}_3\,\mathrm{d}x_1
+\big(h_3-(x_1-k\pi)h_3'\big)\,\mathrm{d}x_2
\nonumber\\
&\qquad
+\mathrm{d}\left(
\frac{\big((x_1-k\pi)v'-v\big)\dot{v}}{h_7}
\right)
\Bigg].\label{eq:RRNiallsol}
\end{align}
%\begin{align}
%\widehat{F}_9&=-c_0\mathrm{vol}_{  \mathrm{AdS}_2} \wedge\mathrm{vol}_{  \mathrm{S}^2} \wedge \mathrm{vol}_{  \mathrm{CY}_2} \wedge \mathrm{d}x_1\left[(x_1-\pi k)(h_3-(x_2-\pi n)\dot{h}_3)\right.\nonumber\\
%&\qquad\qquad\qquad\qquad\quad\qquad\qquad\qquad\left.-\left(\frac{(v-(x_2-n\pi)\dot{v})(v-(x_1-\pi k)v')}{h_7}\right)'\right]\nonumber\\
%&+c_0\mathrm{vol}_{  \mathrm{AdS}_2} \wedge\mathrm{vol}_{  \mathrm{S}^2} \wedge \mathrm{vol}_{  \mathrm{CY}_2} \wedge \mathrm{d}x_2\left[(x_2-\pi n)(h_3-(x_1-\pi k)h_3')\right.\nonumber\\
%&\qquad\qquad\qquad\qquad\qquad\qquad\qquad\quad\left.-\partial_{x_2}\left(\frac{(v-(x_2-n\pi)\dot{v})(v-(x_1-\pi k)v')}{h_7}\right)\right]
%\end{align}
The background above is a supersymmetric solution of Type IIB supergravity provided that, away from localised sources,
\begin{equation}
    h_3''+\ddot{h}_3=0,
    \qquad
    h_7''+\ddot{h}_7=0,
    \qquad
    v''+\ddot{v}=0.
    \label{eq:harmonicConditions}
\end{equation}
The condition on $v$ follows from supersymmetry, while those on $h_3$ and $h_7$ follow from the Bianchi identities, as we show below. The conditions \eqref{eq:harmonicConditions} state that $h_3$, $h_7$ and $v$ are harmonic on the two-dimensional space parametrised by $(x_1,x_2)$, which plays the role of a Riemann surface $\Sigma$. The solutions \eqref{eq:NSNSNiallsol}--\eqref{eq:RRNiallsol} therefore form a family labelled by the choice of these harmonic functions, in the same spirit as the half-BPS AdS$_2\times$S$^2\times\Sigma$ solutions locally asymptotic to AdS$_3\times$S$^3$ constructed in \cite{Chiodaroli:2009yw,Chiodaroli:2009xh}, where the geometry is also fixed by harmonic data on a Riemann surface. The points where $h_3$ and $h_7$ are not harmonic signal localised brane sources, while $v$ controls the AdS$_2$ and S$^2$ warp factors through $\Xi_1$ and $\Xi_2$.

 In what follows we analyse specific choices of these functions and the brane configurations they describe, which lead to different dual quantum mechanics. We restrict to separable configurations,
\begin{equation}
\label{eq:spliting}
h_{3}= h_{1}(x_1)g_{1}(x_2)\,,\qquad h_{7}= h_{2}(x_1)g_2(x_2)\,,\qquad 
v = h_{v}(x_1)\, g_{v}(x_2)\,,
\end{equation}
which keep the problem tractable while still covering a large class of solutions. Each of the functions $h_{1,2}$ and $g_{1,2}$ is taken to be continuous and piecewise linear, with changes of slope at distinguished points along $x_1$ and $x_2$,
\begin{eqnarray}
\label{eq:hs}
\begin{aligned}
h_1 &=
\begin{cases}
\dfrac{\beta_0}{\pi} x_1  \\[2mm]
\alpha_k + \dfrac{\beta_k}{\pi}(x_1 - \pi k)  \\[2mm]
\alpha_{P_1} - \dfrac{\alpha_{P_1}}{\pi}(x_1 - \pi P_1) 
\end{cases}
\quad
h_2 =
\begin{cases}
\dfrac{\nu_0}{\pi} x_1 & \quad 0 \le x_1 \le \pi, \\[2mm]
\mu_k + \dfrac{\nu_k}{\pi}(x_1 - \pi k) & \quad \pi k \le x_1 \le \pi(k+1), \\[2mm]
\mu_{P_{1}} - \dfrac{\mu_{P_{1}}}{\pi}(x_1 - \pi P_1) & \quad \pi P_1 \le x_1 \le \pi(P_1+1),
\end{cases}
\end{aligned}
\end{eqnarray}
\begin{eqnarray}
\label{eq:gs}
\begin{aligned}
g_1 &=
\begin{cases}
\dfrac{\gamma_0}{\pi} x_2  \\[2mm]
\varrho_n+ \dfrac{\gamma_n}{\pi}(x_2 - \pi n)  \\[2mm]
\varrho_{P_2} - \dfrac{\varrho_{P_2}}{\pi}(x_2 - \pi P_2) 
\end{cases}
\quad
g_2 =
\begin{cases}
\dfrac{\theta_0}{\pi} x_2 & \quad 0 \le x_2 \le \pi, \\[2mm]
\sigma_n + \dfrac{\theta_n}{\pi}(x_2 - \pi n) & \quad \pi n \le x_2 \le \pi(n+1), \\[2mm]
\sigma_{P_2} - \dfrac{\sigma_{P_2}}{\pi}(x_2 - \pi P_2) & \quad \pi P_2 \le x_2 \le \pi(P_2+1).
\end{cases}
\end{aligned}
\end{eqnarray}
This structure is the natural analogue of the piecewise-defined functions appearing in the AdS$_3$ parent geometries \cite{Lozano:2019jza, Lozano:2019zvg}.  For a continuous piecewise-linear function the second derivative is a sum of delta functions weighted by the jumps of the first derivative,
\begin{align}
h_1''&=\frac1\pi\sum_{j=1}^{P_1}(\beta_j-\beta_{j-1})\delta(x_1-\pi j),&
h_2''&=\frac1\pi\sum_{j=1}^{P_1}(\nu_j-\nu_{j-1})\delta(x_1-\pi j),\nonumber\\
\ddot g_1&=\frac1\pi\sum_{\ell=1}^{P_2}(\gamma_\ell-\gamma_{\ell-1})\delta(x_2-\pi\ell),&
\ddot g_2&=\frac1\pi\sum_{\ell=1}^{P_2}(\theta_\ell-\theta_{\ell-1})\delta(x_2-\pi\ell),
\label{eq:profileDeltas}
\end{align}
so that, with the separable ansatz,
\begin{equation}
h_3''+\ddot h_3=h_1''g_1+h_1\ddot g_1,
\qquad
h_7''+\ddot h_7=h_2''g_2+h_2\ddot g_2.
\label{eq:separatedLaplacians}
\end{equation}
Within each gauge patch, where $k$ and $n$ are constant, the Bianchi identities of the Page fluxes \eqref{eq:RRNiallsol} read
\begin{align}
\mathrm{d}\widehat{F}_1
&= -c_0\big(h_7''+\ddot{h}_7\big)\,
\mathrm{d}x_1\wedge\mathrm{d}x_2,
\nonumber\\[1ex]
\mathrm{d}\widehat{F}_3
&= c_0\left(-(x_2-n\pi)\big(h_7''+\ddot{h}_7\big)\,
\mathrm{vol}_{\mathrm{AdS}_2}
+(x_1-k\pi)\big(h_7''+\ddot{h}_7\big)\,
\mathrm{vol}_{\mathrm{S}^2}\right)
\wedge\mathrm{d}x_1\wedge\mathrm{d}x_2,
\nonumber\\[1ex]
\mathrm{d}\widehat{F}_5
&= c_0\left((x_1-k\pi)(x_2-n\pi)
\big(h_7''+\ddot{h}_7\big)\,
\mathrm{vol}_{\mathrm{AdS}_2}
\wedge\mathrm{vol}_{\mathrm{S}^2}
+\big(h_3''+\ddot{h}_3\big)\,
\mathrm{vol}_{\mathrm{CY}_2}\right)
\wedge\mathrm{d}x_1\wedge\mathrm{d}x_2,
\nonumber\\[1ex]
\mathrm{d}\widehat{F}_7
&= c_0\left((x_2-n\pi)\big(h_3''+\ddot{h}_3\big)\,
\mathrm{vol}_{\mathrm{AdS}_2}-(x_1-k\pi)\big(h_3''+\ddot{h}_3\big)\,
\mathrm{vol}_{\mathrm{S}^2}
\right)\wedge\mathrm{vol}_{\mathrm{CY}_2}
\wedge\mathrm{d}x_1\wedge\mathrm{d}x_2.
\label{eq:BianchiFT}
\end{align}
All of them are proportional to the combinations in \eqref{eq:separatedLaplacians}. For the profiles \eqref{eq:hs}--\eqref{eq:gs} these vanish inside each interval, while at the interfaces $x_1=\pi j$ and $x_2=\pi\ell$ they are given by the delta functions in \eqref{eq:profileDeltas}, which localise the brane sources. The components along AdS$_2$ are not independent. They are the magnetic ones multiplied by $(x_2-n\pi)$, a factor that arises from the electric component of $B_2$ in $\widehat{F}=e^{-B_2}\wedge F$, so they carry no new information about the sources. At the level of the charges they give the relations between electric and magnetic charges found in Section~\ref{sec:one}, which underlie the baryon-vertex interpretation in Section~\ref{sec:dualQM}. The sources are therefore read off from the magnetic components. Those along S$^2$ measure D1 and D5 charge and carry the factor $(x_1-k\pi)$. Since $(x_1-\pi j)\,\delta(x_1-\pi j)=0$, they have no source term at the $x_1$ interfaces in the patch with $k=j$, and the charges of the D1 and D5 colour branes are carried by the Page fluxes along $x_1$. At the $x_2$ interfaces, instead, the delta functions survive and localise D1 and D5 charge in $x_2$. The components with no such factor, $\mathrm{d}\widehat{F}_1$ and the CY$_2$ component of $\mathrm{d}\widehat{F}_5$, retain all their delta functions and describe the D7 and D3 flavour sources. The two sets are related, since \eqref{eq:BianchiFT} gives
\begin{equation}
\mathrm{d}\widehat{F}_3\big|_{\mathrm{S}^2}=-(x_1-k\pi)\,\mathrm{vol}_{\mathrm{S}^2}\wedge\mathrm{d}\widehat{F}_1~,
\qquad
\mathrm{d}\widehat{F}_7\big|_{\mathrm{S}^2}=-(x_1-k\pi)\,\mathrm{vol}_{\mathrm{S}^2}\wedge\mathrm{d}\widehat{F}_5\big|_{\mathrm{CY}_2}~,
\label{eq:inducedD1D5}
\end{equation}
where $|_{\mathrm{S}^2}$ and $|_{\mathrm{CY}_2}$ denote the components along the corresponding volume forms. As for the components along AdS$_2$, the factor $(x_1-k\pi)$ arises from the magnetic component of $B_{2}$ through  $\widehat{F}=e^{-B_2}\wedge F$.
The D1 and D5 sources at the $x_2$ interfaces are therefore not independent. The D3 and D7 branes located there wrap the S$^2$ and carry, respectively, D1 and D5 charge induced by $B_2$, with the corresponding density fixed by \eqref{eq:inducedD1D5}.

The brane configuration, including the NS5-branes and fundamental
strings, is summarised in Table~\ref{table:general}.
Configurations sharing the directions indicated in the table
are grouped in the same row, while their different source
distributions are distinguished by the Page fluxes and their
Bianchi identities. We will determine the corresponding
charges and discuss their roles in the dual quantum mechanics for
the specific backgrounds considered in the following sections.

\begin{table}[t]
\centering
\renewcommand{\arraystretch}{1.2}
\begin{tabular}{c|| c|cccc|c|ccc|c}
\hline\hline
 Brane & $t$ & $y_1$ & $y_2$ & $y_3$ & $y_4$ & $x_1$ & $z$ & $\vartheta$ & $\varphi$ & $x_2$ \\\hline\hline
D1  & $\bullet$ & - & - & - & - & $\bullet$ & - & - & -& - \\
D3  & $\bullet$ & - & - & - & - & - & $\bullet$ & $\bullet$ & $\bullet$ & -\\
D5  & $\bullet$ & $\bullet$ & $\bullet$ & $\bullet$ & $\bullet$ & $\bullet$ & - & - & - & - \\
D7  & $\bullet$ & $\bullet$ & $\bullet$ & $\bullet$ &  $\bullet$&  -&  $\bullet$& $\bullet$ & $\bullet$ & - \\
NS5 & $\bullet$ & $\bullet$ &  $\bullet$& $\bullet$ & $\bullet$ & - & - & - & - & $\bullet$ \\
F1 & $\bullet$ & - & - & - & - & - & - & - & - & $\bullet$ \\
\hline\hline
\end{tabular}
\caption{BPS brane intersection underlying the geometry in \eqref{eq:NSNSNiallsol}--\eqref{eq:RRNiallsol}. The quantum mechanics lives along the time direction $t$. The directions $(y_{1},\ldots,y_{4})$ span the CY$_{2}$ wrapped by the D5 and D7 branes, and $x_{1}$ and $x_{2}$ parametrise the Riemann surface $\Sigma$. The coordinates $(z,\vartheta,\varphi)$ are spherical coordinates on the $\mathbb{R}^{3}$ transverse to the D1, D5, NS5 and F1. In the near-horizon limit $z$ and $t$ combine into AdS$_{2}$, and $(\vartheta,\varphi)$ parametrise the S$^{2}$, on which the SO$(3)$ symmetry acts.}\label{table:general}
\end{table}

\vspace{0.5cm} 
Using \eqref{eq:ccholAdS2}, the holographic central charge of this family reads
\begin{equation}
\label{eq:ccholfamily}
    c_{\text{hol}} \= \frac{6\,c_0^2}{\pi^2}\int \Xi_1\;dx_1 \,dx_2 \= \frac{6\,c_0^2}{\pi^2}\int \big(h_1h_2g_1g_2-h_v^2\,\dot{g}_v^{\,2}\big)\,dx_1 \,dx_2~,
\end{equation}
where we have set $\mathrm{Vol}(\mathrm{CY}_2)=16\pi^4$. Since this expression only depends on $\Xi_1$, it can be computed interval by interval for the piecewise-linear profiles \eqref{eq:hs}--\eqref{eq:gs}. We shall compare it with the central charge of the dual quantum mechanics in Section~\ref{sec:dualQM}.

\subsection{U(1) isometry along $\partial_{x_2}$}
\label{subsec:x2asanisometry}

In this subsection we consider the case in which $\partial_{x_2}$ generates a $U(1)$ isometry. As observed in \cite{Legramandi:2023fjr}, for suitable choices of the functions $(h_1,h_2,h_v,g_1,g_2,g_v)$ the backgrounds \eqref{eq:NSNSNiallsol}--\eqref{eq:RRNiallsol} then reproduce those obtained and studied in detail in \cite{Lozano:2020txg}.\footnote{These geometries were obtained by Abelian T-duality along the Hopf fibre of the AdS$_3$ in the parent geometry \cite{Lozano:2019emq}.} In particular, by choosing
\begin{align}
&h_1=h_4~,\qquad h_2=h_8~,\qquad h_v=u~,\qquad g_1=g_2=4g_v=\frac{1}{c_0}~,\nonumber\\
&x_1=\frac{\rho}{2}~,\qquad x_2=\frac{\psi}{2}~,
\end{align}
the background reduces to eqs.~(3.2)--(3.3) of \cite{Lozano:2020txg}. Since $\psi\in[0,2\pi]$, the coordinate $x_2$ now ranges over the single interval $[0,\pi]$. More generally, we take
\begin{equation}
g_1=\varrho~,\qquad g_2=\sigma~,\qquad g_v=c_v~,
\end{equation}
with $\varrho$, $\sigma$ and $c_v$ constants, so that the background depends non-trivially only on $x_1$. The solutions of \cite{Lozano:2020txg} correspond to $\varrho=\sigma=4c_v=1/c_0$. 

The Page fluxes and Bianchi identities follow from \eqref{eq:RRNiallsol} and \eqref{eq:BianchiFT} with $\dot g_1=\dot g_2=\dot g_v=0$. Only the delta functions at the $x_1$ interfaces survive, and one recovers the brane content of \cite{Lozano:2020txg}: D1 and D5 colour branes dissolved in the fluxes, and D3 and D7 flavour branes localised at $x_1=\pi k$. Since $x_2\in[0,\pi]$, no electric large gauge transformation is needed and we can set $n=0$. Using \eqref{eq:magneticPage}, the charges in the interval $[\pi k,\pi(k+1)]$ are
\begin{align}
Q_{1}^{[k]}
&=
\frac{1}{(2\pi)^6}
\int \widehat{F}_7
=
c_0\,\varrho\,\alpha_k,
&
Q_{5}^{[k]}
&=
\frac{1}{(2\pi)^2}
\int \widehat{F}_3
=
c_0\,\sigma\,\mu_k,
\nonumber\\[2mm]
Q_{3}^{[k]}
&=
\frac{1}{(2\pi)^4}
\int d\widehat{F}_5
=
c_0\,\varrho\,
\bigl(\beta_{k-1}-\beta_k\bigr),
&
Q_{7}^{[k]}
&=
\int d\widehat{F}_1
=c_0\,\sigma\,
\bigl(\nu_{k-1}-\nu_k\bigr),
\label{eq:charges_X2isometry}
\end{align}
where the orientation has been chosen so that the flavour charges are positive for the convex profiles \eqref{eq:hs}. For $\varrho=\sigma=1/c_0$ these are the charges in eq.~(3.9) of \cite{Lozano:2020txg}, and the brane configuration is the one in Table~\ref{table:general}.

Specialising \eqref{eq:ccholfamily} to these profiles, the holographic central charge becomes
\begin{align}
    c_{\text{hol}}&= \frac{6c_0^2}{\pi^2}\,\varrho\;\sigma\int h_1h_2 \;dx_1\, dx_2\nonumber\\
    &=c_0^2\,\varrho\;\sigma\sum_{k=0}^{P_1}\big(6\alpha_k\mu_k+3\alpha_k\nu_k+3\beta_k\mu_k+2\beta_k\nu_k\big)~,
    \label{eq:holographicCentralChargeX2}
\end{align}
where, following \eqref{eq:hs}, $\alpha_0=\mu_0=0$, $\beta_{P_1}=-\alpha_{P_1}$ and $\nu_{P_1}=-\mu_{P_1}$. For $\varrho=\sigma=1/c_0$ this agrees with eq.~(4.4) of \cite{Lozano:2020txg}. The first term is controlled by the D1- and D5-brane charges, whereas the remaining terms depend on the data associated with the flavour sector.

\vspace{0.5cm}
Let us also comment on the sector in which $\partial_{x_1}$ generates a $U(1)$ isometry. As pointed out in \cite{Legramandi:2023fjr}, it contains the backgrounds of \cite{Lozano:2021rmk}. By choosing
\begin{align}
    &g_1=h_4~,\qquad g_2=h_8~,\qquad g_v=u~,\qquad
    h_1=h_2=4h_v=\frac{1}{c_0}~,\nonumber\\
    &x_1=\frac{\rho}{2}~,\qquad x_2=\frac{\psi}{2}~,
\end{align}
and exchanging
\begin{equation}
\label{eq:exchange}
    \psi \longleftrightarrow \rho\,,
\end{equation}
the backgrounds \eqref{eq:NSNSNiallsol}--\eqref{eq:RRNiallsol} reproduce eqs.~(2.1)--(2.2) of \cite{Lozano:2021rmk}. The exchange is needed because those solutions are related to those of \cite{Lozano:2020txg} by a double analytic continuation \cite{Lozano:2021rmk}.

\subsection{Type A embedding and global completion}
\label{sec:FTtypeI}

In this subsection we show that the Type A backgrounds of Section~\ref{sec:one} belong to the class \eqref{eq:NSNSNiallsol}--\eqref{eq:RRNiallsol}, and we use this embedding to propose a global completion along $x_2$. Upon choosing
\begin{align}
\label{eq:transformation_TypeI}
&h_1=h_4, \qquad h_2=h_8,\qquad h_v=u,\nonumber\\ &g_1=\frac{r_1}{4c_0}~,\qquad g_2=\frac{r_1}{4c_0},~\qquad g_v=\frac{r_1}{2^4c_0}~, \nonumber\\ &x_1=\frac{\rho}{2}~,\qquad x_2=\frac{r_1}{2}~,
\end{align}
the NSNS sector reduces to \eqref{eq:typeINSNS} and the RR fluxes to \eqref{eq:typeIRR}. The Type A backgrounds thus correspond to linear profiles along $x_2$, with $g_1=g_2=4g_v$. In what follows we take $h_1$ and $h_2$ to be the piecewise-linear profiles \eqref{eq:hs}, and $h_v=u_0\,x_1$.

Following \cite{Ramirez:2021tkd}, we complete the profile along $x_2$ by gluing the NATD solution to its mirror image across a turning point at $x_2=\pi P_2$,
\begin{equation}
\label{eq:completionTypeI}
g_1(x_2)=g_2(x_2)=4g_v(x_2)=\frac{1}{2c_0}
\begin{cases}
x_2
& \quad \pi\le x_2\le \pi P_2,
\\[2mm]
2\pi P_2-x_2
& \quad \pi P_2\le x_2\le \pi(2P_2-1).
\end{cases}
\end{equation}
The geometry is well defined for $\Xi_1>0$, is equivalent to the lower bound \eqref{eq:r0natd} of the Type A solution, which is written in terms of the separated functions. Since $\dot g_v^{\,2}=1/(64c_0^2)$ on both branches, this condition becomes
\begin{equation}
2c_0\,g_1(x_2)>x_2^{(*)}(x_1)~,
\qquad
x_2^{(*)}(x_1)=\frac{h_v}{4\sqrt{h_1h_2}}~.
\label{eq:x2starTypeI}
\end{equation}
On the first interval of \eqref{eq:hs} the functions $h_1$ and $h_2$ are proportional to $x_1$ and the bound is constant. We fix it to $x_2^{(*)}=\pi$, which amounts to $u_0=4\sqrt{\beta_0\nu_0}$ and matches the lower end of \eqref{eq:completionTypeI}. On the interval $k$ one has $h_1=(\alpha_k-k\beta_k)+\beta_k x_1/\pi$, with $\alpha_k-k\beta_k=\sum_{j<k}(\beta_j-\beta_k)$, and similarly for $h_2$. The bound then stays equal to $\pi$ as long as there are no flavour branes to the left, and grows once they are crossed, while remaining finite. On the last interval, $x_2^{(*)}(x_1)$ grows without bound as $x_1$ approaches $\pi(P_1+1)$. For finite $P_2$, however, the physical domain terminates earlier, where $x_2^{(*)}(x_1)=\pi P_2$. The endpoint $x_1=\pi(P_1+1)$ of the prescribed profiles therefore lies outside the physical domain.

With the completion \eqref{eq:completionTypeI}, the condition \eqref{eq:x2starTypeI} restricts the physical domain to
\begin{equation}
\label{eq:domainTypeI}
x_2^{(*)}(x_1)<x_2<2\pi P_2-x_2^{(*)}(x_1)~,
\qquad
x_2^{(*)}(x_1)<\pi P_2~.
\end{equation}
We work at large $P_2$, keeping the profiles along $x_1$ fixed. To compute the charges and the holographic central charge at leading order, we approximate this domain by the rectangle $0\leq x_1\leq\pi(P_1+1)$, $\pi\leq x_2\leq\pi(2P_2-1)$, since the excluded regions are strips whose contributions are subleading in this limit. At the boundaries of the physical domain, where $\Xi_1=0$, the geometry behaves as near the lower bound of Section~\ref{sec:one},  which we interpreted there as a smeared OF1-plane. We regard \eqref{eq:completionTypeI} as a proposal for the global completion, whose leading holographic central charge we compare with the dual quantum mechanics in Section~\ref{sec:dualQM}. A complete global analysis would also require a more detailed study of the turning point.%\footnote{The reflection of $g_v$ introduces a kink in $v$, so that the condition on $v$ in \eqref{eq:harmonicConditions} holds on each branch but not at the turning point, where $v''+\ddot v=-\frac{u_0\,x_1}{4c_0}\,\delta(x_2-\pi P_2)$. Keeping $g_v$ linear on both branches would avoid this, but the Page charges measured along $x_1$ would then depend on $x_2$ on the second branch. Whether this localised term can be accounted for by the sources at the turning point, consistently with supersymmetry, remains to be established.}

Along $x_2$ we integrate the charges \eqref{eq:DpmagneticchargestypeI} over the electric gauge patches of Section~\ref{sec:one}, whose boundaries approach $x_2=\pi n$ away from the ends of the $x_2$ interval. This introduces the factor

\begin{equation}
f_n=
\begin{cases}
2n+1,
& 1\leq n\leq P_2-1,
\\[2mm]
4P_2-(2n+1),
& P_2\leq n\leq 2P_2-2,
\end{cases}
\label{eq:fnTypeI}
\end{equation}
and the charges in the cell $[k,n]$ read
\begin{align}
Q_{1}^{[k,n]} &= \frac{\pi}{4} \alpha_k f_n,
&
Q_{5}^{[k,n]} &= -\frac{\pi}{4} \mu_k f_n,
\nonumber\\[2mm]
Q_{3}^{[k,n]} &= \frac{\pi}{4}(\beta_{k-1}-\beta_k) f_n,
&
Q_{7}^{[k,n]} &= -\frac{\pi}{4}(\nu_{k-1}-\nu_k) f_n,
\label{eq:chargesTypeI}
\end{align}
where the signs depend on the orientation of the Page-charge integrals and the ranks are given by their absolute values. The dependence on $x_1$ is encoded in $(\alpha_k,\mu_k,\beta_k,\nu_k)$ whereas the dependence on $x_2$ in $f_n$. The D3 and D7 sources remain localised at the interfaces $x_1=\pi k$, and the factor $f_n$ only reflects the integration of their charge density over the corresponding $x_2$ interval.

The kinks of $g_1$ and $g_2$ at $x_2=\pi P_2$ are sources of primed branes. According to \eqref{eq:BianchiFT}, the relations $\ddot g_1=\ddot g_2=-\frac{1}{c_0}\delta(x_2-\pi P_2)$ describe D7$'$ and D3$'$ branes smeared along $x_1$ with densities $h_2$ and $h_1$, respectively. These branes also carry D5$'$ and D1$'$ charge, with densities $(x_1-k\pi)h_2$ and $(x_1-k\pi)h_1$, induced by $B_2$ as discussed around \eqref{eq:inducedD1D5}. Only the D7$'$ and D3$'$ charges are therefore independent. They are the ones related to $Q^e_1$ and $Q^e_5$ through \eqref{eq:DpelectricmagneticchargestypeI}, the relations used in Section~\ref{sec:dualQM}. The relations involving D5$'$ and D1$'$ add no new information. 

Substituting the completed profiles into \eqref{eq:ccholfamily} and integrating over the rectangle, one finds
\begin{align}
c_{\rm hol}
&=
\sum_k
\int_{\pi k}^{\pi(k+1)}
\Bigg[
\int_{\pi}^{\pi P_2}
\frac{
48x_2^2 h_1h_2-3h_v^2
}{
32\pi^2
}\,dx_2
\nonumber\\[-1mm]
&\hspace{3.2cm}
+
\int_{\pi P_2}^{\pi (2P_2-1)}
\frac{
48\left(2\pi P_2-x_2\right)^2h_1h_2
-3h_v^2
}{
32\pi^2
}\,dx_2
\Bigg]dx_1,\nonumber\\
&=
\pi^2\sum_k
\Bigg[
\left(
\alpha_k\mu_k
+\frac{1}{2}\beta_k\mu_k
+\frac{1}{2}\alpha_k\nu_k
+\frac{1}{3}\beta_k\nu_k
\right)
\left(P_2^3-1\right)-\frac{u_0^2}{16}
\left(P_2-1\right)
\left(3k^2+3k+1\right)
\Bigg].
\label{eq:cholTypeI}
\end{align}
The second term, controlled by $v$, is suppressed by $P_2^{-2}$ with respect to the first. As for the charges, we only keep the leading term at large $P_2$, which is the one we compare with the dual quantum mechanics in Section~\ref{sec:dualQM}.

Finally, the class \eqref{eq:NSNSNiallsol}--\eqref{eq:RRNiallsol} also contains the double non-Abelian T-dual of AdS$_3\times$S$^3\times$CY$_2$, obtained by dualising an SL$(2,\mathbb{R})$ isometry of AdS$_3$ and an SU$(2)$ isometry of S$^3$. It belongs to the Type A family, with profiles that are linear in both $x_1$ and $x_2$,
\begin{align}
&h_1=2L^2\mathscr{M}^4\,x_1~,\qquad h_2=2L^2\,x_1~,\qquad h_v=8L^4\mathscr{M}^2\,x_1~,\nonumber\\
&g_1=\frac{x_2}{2c_0}~,\qquad g_2=\frac{x_2}{2c_0}~,\qquad g_v=\frac{x_2}{2^3c_0}~,
\label{eq:doubleNATDAdS3S3T4}
\end{align}
where $L$ is the radius of AdS$_3$ and S$^3$ and $\mathscr{M}$ that of the CY$_2$ in the original solution. The functions $h_3$, $h_7$ and $v$ are bilinear in $x_1$ and $x_2$, and therefore harmonic, and the NSNS sector \eqref{eq:NSNSNiallsol} becomes
\begin{align}
ds^2 =&\frac{L^2 x_2^2}{x_2^2-L^4} ds^2_{\mathrm{AdS}_2}+\frac{L^2x_1^2}{x_1^2+L^4}
ds^2_{\mathrm{S}^2}+\mathscr{M}^2 ds^2_{\mathrm{CY}_2}+\frac{dx_1^2+dx_2^2}{L^2},
\nonumber\\
e^{-2\Phi}=&
L^4(x_2^2-L^4)(x_1^2+L^4),\qquad
B_2=
-\frac{x_2^3}{x_2^2-L^4}\mathrm{vol}_{\mathrm{AdS}_2}
+\frac{x_1^3}{x_1^2+L^4}\mathrm{vol}_{\mathrm{S}^2}.
\label{eq:doubleNATDbackground}
\end{align}
Along $x_2$ the profiles are those of the Type A solution in \eqref{eq:transformation_TypeI}. Along $x_1$, however, $h_1$, $h_2$ and $h_v$ grow linearly without bound, unlike the profiles \eqref{eq:hs}, which already close at both ends, so we complete the background along both directions. Along $x_2$ we take the profiles \eqref{eq:completionTypeI}, with the ends $x_2=\pi$ and $x_2=\pi(2P_2-1)$ replaced by $L^2$ and $2\pi P_2-L^2$, and along $x_1$ we glue the solution to its mirror image at $x_1=\pi P_1$,
\begin{equation}
h_1=\mathscr{M}^4h_2=2L^2\mathscr{M}^4
\begin{cases}
x_1 & \quad 0\le x_1\le\pi P_1~,\\[1mm]
2\pi P_1-x_1 & \quad \pi P_1\le x_1\le 2\pi P_1~.
\end{cases}
\label{eq:hsDoubleNATD}
\end{equation}
This profile is of the form \eqref{eq:hs} so the completed background is a Type A solution. The function $h_v$ is not reflected and stays linear, while $g_v$ is reflected as in the completion \eqref{eq:completionTypeI}. The lines $x_1=\pi P_1$ and $x_2=\pi P_2$ divide the domain into four sectors. In the first one, with $0\le x_1\le\pi P_1$ and $L^2\le x_2\le\pi P_2$, the background is given by  \eqref{eq:doubleNATDbackground}, and in the other three sectors it is obtained by reflecting $h_{1,2}$, $g_{1,2,v}$ or both. The kinks along $x_1=\pi P_1$ correspond to D3 and D7 flavour branes, smeared along $x_2$ with densities $g_1$ and $g_2$, and those along $x_2=\pi P_2$ correspond to the D3$'$ and D7$'$ branes discussed above, smeared along $x_1$ with densities $h_1$ and $h_2$.

The main difference with the profiles \eqref{eq:hs} is in the domain. With these profiles the bound \eqref{eq:x2starTypeI} is $x_2^{(*)}=L^2$ on the first half of the $x_1$ interval, which is why the first branch along $x_2$ starts at $x_2=L^2$. On the second half the bound grows as $L^2x_1/(2\pi P_1-x_1)$, which diverges at $x_1=2\pi P_1$ and already equals $(2P_1-1)L^2$ at the beginning of the last interval. The physical domain is again given by \eqref{eq:domainTypeI}, and we approximate it by the rectangle $0\le x_1\le 2\pi P_1$, $L^2\le x_2\le 2\pi P_2-L^2$. This approximation now requires $P_2\gg P_1L^2$, rather than $P_2\gg1$.

For CY$_2=$T$^4$ this background arises, after an analytic continuation, as a limit of the Type IIB embedding of the $\lambda$-deformed SL$(2,\mathbb{R})\times$SU$(2)$ model \cite{Sfetsos:2014cea,Itsios:2023kma}. The $\lambda$-deformation requires AdS$_3$ and S$^3$ to be unwarped group manifolds, while non-Abelian T-duality only uses their isometries and therefore also applies to backgrounds in which they are warped over other directions, such as the parent solution of Appendix~\ref{app:B-Appendix}. As for other non-Abelian T-duals, the dualisation only determines this background locally and does not fix the ranges of $x_1$ and $x_2$. The completions \eqref{eq:completionTypeI} and \eqref{eq:hsDoubleNATD} provide a global definition of the background, which makes its brane content manifest and allows us to propose its dual quantum mechanics in terms of quivers in Section~\ref{sec:dualQM}.
%As mentioned in Section~\ref{sec:twoandthree}, for CY$_2=$T$^4$ this background arises, after an analytic continuation, as a limit of the Type IIB embedding of the $\lambda$-deformed SL$(2,\mathbb{R})\times$SU$(2)$ model \cite{Sfetsos:2014cea,Itsios:2023kma}. As for other non-Abelian T-duals, the dualisation only determines it locally and does not fix the ranges of $x_1$ and $x_2$. The completions \eqref{eq:completionTypeI} and \eqref{eq:hsDoubleNATD} provide a global definition of the background, which makes its brane content manifest and allows us to propose its dual quantum mechanics in terms of quivers in Section~\ref{sec:dualQM}.

\section{Dual quantum mechanics}
\label{sec:dualQM}

In this section we propose the quantum mechanics dual to the Type A backgrounds of Section~\ref{sec:FTtypeI}. We start from the quantum mechanics of the U(1)-invariant backgrounds, reviewed in Section~\ref{subsec:qmUone} and identified in \cite{Lozano:2020txg}. In Section~\ref{subsec:qmTypeI} we build the dual of the completed solution from copies of this quiver along $x_2$, with the ranks determined by $f_n$, compare its central charge with \eqref{eq:cholTypeI} and identify the baryon vertices. We close with some comments on the general solution in Section~\ref{subsec:qmGeneral}.

\subsection{Quantum mechanics of the U(1)-invariant backgrounds}
\label{subsec:qmUone}

The ${\cal N}=4$ quantum mechanics dual to the backgrounds of Section~\ref{subsec:x2asanisometry} was studied in \cite{Lozano:2020txg}, where it was obtained by dimensional reduction of the two-dimensional ${\cal N}=(0,4)$ quiver theory dual to the parent AdS$_3$ solution \cite{Lozano:2019emq,Lozano:2019zvg}. We only review the elements needed for the Type A analysis of Section~\ref{subsec:qmTypeI}.

When AdS$_3$ is written as a Hopf fibration over AdS$_2$, its boundary consists of two null cylinders. Therefore, the two-dimensional theory is discrete light-cone quantised and only one of the two SL$(2,\mathbb{R})$ factors survives \cite{Lozano:2020txg}. The quantum mechanics follows from reducing the two-dimensional theory along its spatial direction, keeping the time dependence and the zero modes. This preserves the four right-moving supercharges of the ${\cal N}=(0,4)$ theory, giving an ${\cal N}=4$ quantum mechanics. The reduction does not change the gauge groups or the representations of the matter fields, so the quiver is inherited from the two-dimensional theory.  At the level of the branes, the Abelian T-duality along the Hopf fibre maps
\begin{equation}
\mathrm{D2}\longrightarrow\mathrm{D1},
\qquad
\mathrm{D6}\longrightarrow\mathrm{D5},
\qquad
\mathrm{D4}\longrightarrow\mathrm{D3},
\qquad
\mathrm{D8}\longrightarrow\mathrm{D7}.
\end{equation}

\begin{figure}[t]
\centering
\includegraphics[width=0.85\textwidth]{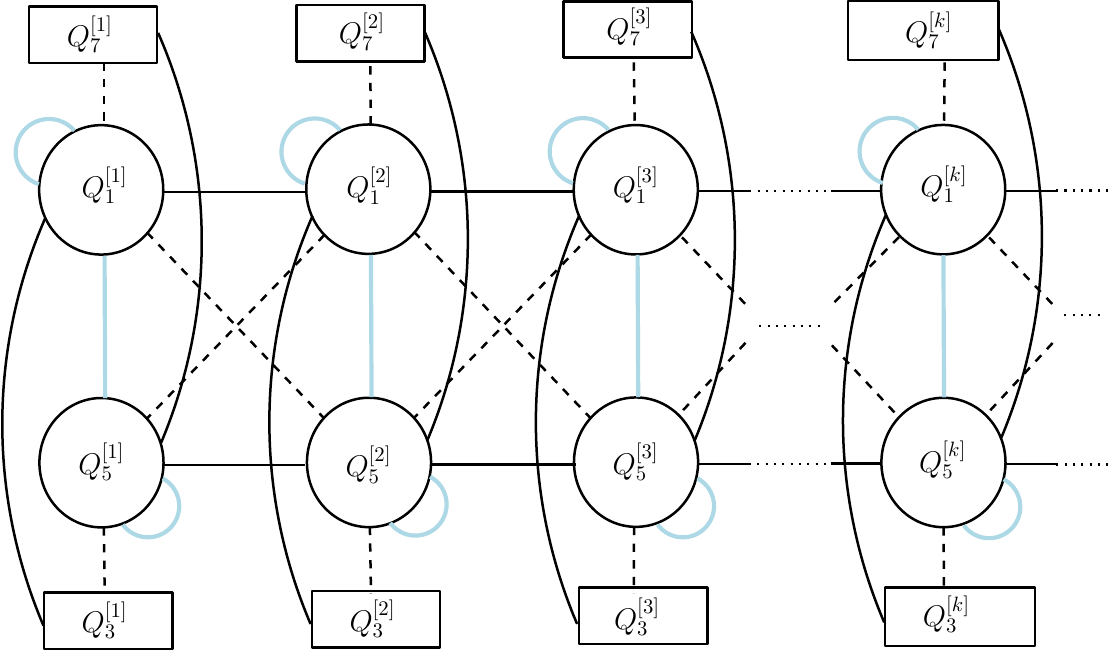}
\caption{Quiver of the ${\cal N}=4$ quantum mechanics dual to the background invariant under the $\mathrm{U}(1)$ generated by $\partial_{x_2}$. Circles denote gauge nodes and squares flavour groups. The light-blue loops are the ${\cal N}=4$ adjoint hypermultiplets, and the light-blue vertical lines the bifundamental hypermultiplets between the two nodes in the same interval. Solid horizontal lines are the ${\cal N}=8$ twisted hypermultiplets between adjacent D1 nodes and between adjacent D5 nodes, and dashed diagonal lines the ${\cal N}=2$ Fermi multiplets between D1 and D5 nodes in neighbouring intervals. Dashed vertical lines connect each node to its adjacent flavour group through Fermi multiplets, and solid curved lines to the opposite flavour group through twisted hypermultiplets.}\label{fig:quiverX1}
\end{figure}

The D1 and D5 colour branes give the gauge nodes, with gauge group ${\rm U}(Q_1^{[k]})\times{\rm U}(Q_5^{[k]})$ in the $k$-th interval, and the D7 and D3 flavour branes give flavour groups ${\rm SU}(Q_7^{[k]})$ and ${\rm SU}(Q_3^{[k]})$, attached to the D1 and D5 colour groups, with the charges given in \eqref{eq:charges_X2isometry}. Each gauge node carries a vector multiplet and a hypermultiplet in the adjoint, descending from ${\cal N}=(0,4)$ multiplets. Adjacent D1 nodes, and adjacent D5 nodes, are connected by bifundamental fields descending from ${\cal N}=(4,4)$ twisted hypermultiplets, and the two nodes in the same interval by a bifundamental descending from an ${\cal N}=(0,4)$ hypermultiplet. D1 and D5 nodes in neighbouring intervals are connected by ${\cal N}=(0,2)$ Fermi multiplets. Each gauge node couples to its adjacent flavour group, D1--D7 and D5--D3, through a fundamental ${\cal N}=(0,2)$ Fermi multiplet, and to the opposite flavour group, D1--D3 and D5--D7, through an ${\cal N}=(4,4)$ twisted hypermultiplet. The two-dimensional labels only indicate the origin of the fields, and should not be read as chiral supersymmetry labels in one dimension. Although there is no gauge anomaly in one dimension, the flavour ranks are fixed by the anomaly cancellation of the two-dimensional theory, and they agree with the D3 and D7 charges in \eqref{eq:charges_X2isometry}. The proposal in \cite{Lozano:2020txg} is that the quiver in Figure~\ref{fig:quiverX1} describes the ultraviolet ${\cal N}=4$ quantum mechanics, and its infrared fixed point is conjectured to be the superconformal quantum mechanics dual to the AdS$_2$ background.

Since the quantum mechanics descends from a two-dimensional ${\cal N}=(0,4)$ theory, its central charge can be computed as in two dimensions \cite{Lozano:2020txg,Lozano:2020bxo,Lozano:2021rmk,Ramirez:2021tkd,Lozano:2020sae},
\begin{equation}
c_{qm}
=
6\left(n_{\mathrm{hyp}}-n_{\mathrm{vec}}\right),
\label{eq:fieldTheoryCentralChargeX2}
\end{equation}
where only the untwisted hypermultiplets and the vector multiplets contribute, since the fermions in the twisted hypermultiplets and in the Fermi multiplets have vanishing charge under the U$(1)_R$ inherited from two dimensions. This prescription allows us to compute the central charge $c_{qm}$ from the quiver. In the two-dimensional theory the central charge is fixed by the anomaly of the superconformal U$(1)_R$, and by 't Hooft anomaly matching it can be evaluated with the field content of the ultraviolet quiver, where only fermions charged under U$(1)_R$ contribute. There are no such anomalies in one dimension, so $c_{qm}$ is the quantity inherited from the two-dimensional theory. For the quiver in Figure~\ref{fig:quiverX1},
\begin{equation}
n_{\rm hyp}=\sum_k \Big(Q_1^{[k]}Q_5^{[k]}+\big(Q_1^{[k]}\big)^2+\big(Q_5^{[k]}\big)^2\Big)~,\qquad n_{\rm vec}=\sum_k \Big(\big(Q_1^{[k]}\big)^2+\big(Q_5^{[k]}\big)^2\Big)~,
\end{equation}
so the adjoint contributions cancel and
\begin{equation}
c_{qm}=6\sum_k Q_1^{[k]}Q_5^{[k]}=6\,c_0^2\,\varrho\sigma \sum_k \alpha_k\mu_k~.
\label{eq:quantumMechanicalCentralChargeX2}
\end{equation}
The sum over $k$ is finite because the quiver terminates at both ends. The profiles \eqref{eq:hs} vanish at $x_1=0$ and $x_1=\pi(P_1+1)$, so the ranks vanish in the first interval, $\alpha_0=\mu_0=0$, and the last gauge nodes, in the interval $k=P_1$, are closed by the flavour groups at the endpoint of the quiver $x_1=\pi P_1$, whose ranks are fixed by anomaly cancellation in the two-dimensional theory. Without this completion the quiver would be infinite and $c_{qm}$ would diverge. This result reproduces the first term of the holographic result \eqref{eq:holographicCentralChargeX2}. The remaining terms, $c_0^2\varrho\sigma\sum_k(3\alpha_k\nu_k+3\beta_k\mu_k+2\beta_k\nu_k)$, are subleading for long quivers with large ranks, since $\alpha_k$ and $\mu_k$ grow linearly with $k$. We shall refer to this regime, $P_1\gg1$, as the holographic limit. In this limit the two central charges agree at leading order, as found in \cite{Lozano:2020txg}.

\subsection{Quantum mechanics of the Type A backgrounds}
\label{subsec:qmTypeI}
The charges \eqref{eq:chargesTypeI} carry a double label $[k,n]$, so the brane configuration is organised in cells over the $(x_1,x_2)$ plane. A similar situation was analysed in \cite{Conti:2025djz} for a background obtained by NATD. There, the field theory was built from the quiver of the Abelian T-dual background, with the additional direction generated by the duality associated with fundamental strings, Wilson lines and baryon vertices. We build the quantum mechanics dual to the Type A backgrounds in the same way, starting from the quiver of Section~\ref{subsec:qmUone} and treating $x_2$ as the direction generated by the duality.

\begin{figure}[t]
\centering
\includegraphics[width=0.9\textwidth]{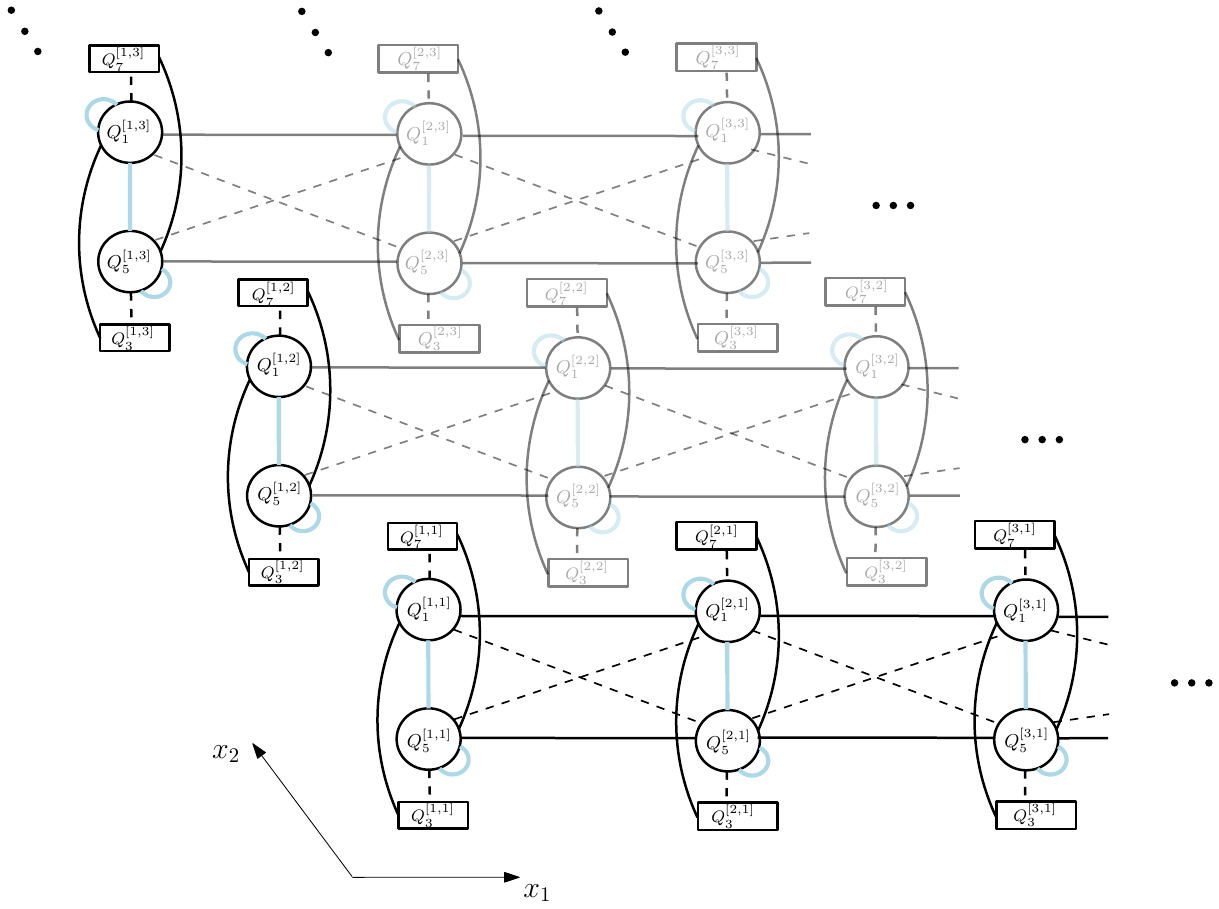}
\caption{Schematic organisation of the charge data over the $(x_1,x_2)$ plane. Each cell $[k,n]$ carries colour ranks $Q_{1,5}^{[k,n]}$, and flavour multiplicities $Q_{3,7}^{[k,n]}$ at the $x_1$ interfaces. Each row is the linear quiver along $x_1$ at fixed $n$, with ranks and multiplicities rescaled by $f_n$.}\label{fig:quivercell}
\end{figure}

At fixed $n$ the factor $f_n$ is constant, and the dependence on $k$ comes only from the profiles along $x_1$. The charges of such a slice therefore reproduce the linear quiver of Figure~\ref{fig:quiverX1}, with colour ranks and flavour multiplicities rescaled by $f_n$, and the matter content is inherited from the quantum mechanics of Section~\ref{subsec:qmUone}. The flavour branes remain at the interfaces $x_1=\pi k$, so all slices have the same quiver topology. We propose that the quantum mechanics dual to the completed geometry is the collection of these slices for $1\le n\le 2P_2-2$, as represented in Figure~\ref{fig:quivercell}. The figure organises the charge data across the $(x_1,x_2)$ plane, and the linear quivers at different values of $n$ are not connected to each other: there is no second set of NS5-branes along $x_2$, and no matter multiplets connect neighbouring slices. Global quantities are then obtained by summing over $n$, and we test this proposal below with the central charge.

\begin{figure}[t]
\centering
\includegraphics[width=1\textwidth]{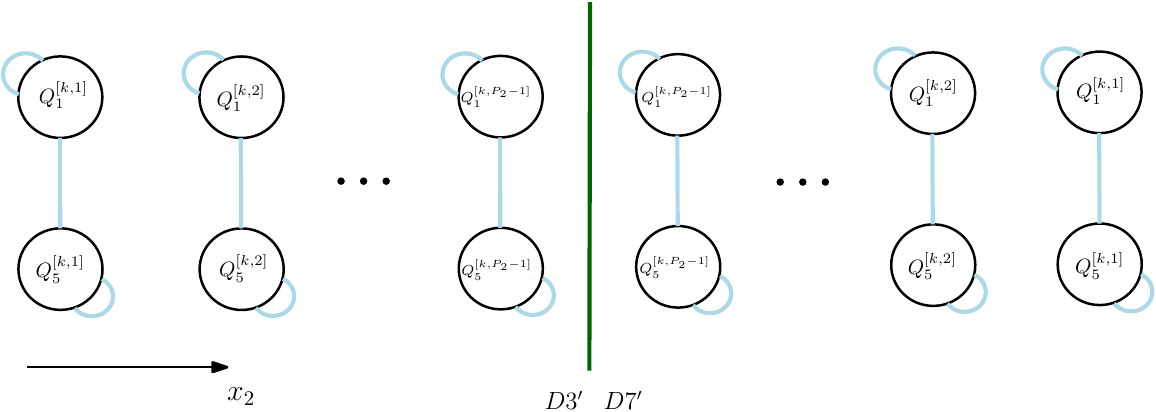}
\caption{Schematic distribution of the colour charges along the $x_2$ direction at fixed $k$. Each pair of nodes carries the D1- and D5-brane charges, which are proportional to $f_n$ in \eqref{eq:fnTypeI}. They increase along the first branch of the completion, reach their maximum in the two central intervals and decrease symmetrically along the second branch. Since $f_{2P_2-1-n}=f_n$, the pairs on the second branch carry the same charges as their mirror images on the first branch, and are labelled accordingly. The vertical line marks the turning point, where the D3$'$ and D7$'$ branes sit.}
\label{fig:quivertypeI}
\end{figure}

To make the structure of Figure~\ref{fig:quivercell} more transparent, consider a slice at fixed $k$. Along this slice the colour charges depend on $n$ only through $f_n$: they increase along the first branch of the completion, reach their maximum in the two central intervals and decrease along the second branch, as displayed in Figure~\ref{fig:quivertypeI}. Their change of slope occurs at the turning point, where the D3$'$ and D7$'$ branes are located. In the slice-wise interpretation adopted here these branes do not give rise to flavour groups of a quiver along $x_2$. They enter the quantum mechanics through the fundamental strings ending on the colour branes as discussed at the end of this section.

As in Section~\ref{subsec:qmUone}, only the untwisted hypermultiplets and the vector multiplets contribute to $c_{qm}=6(n_{hyp}-n_{vec})$, and the adjoint contributions cancel slice by slice. Summing over the slices and using the reflection symmetry $f_{2P_2-1-n}=f_n$, one finds
\begin{align}
c_{qm}
&=6\sum_k\sum_{n=1}^{2P_2-2}Q_{1}^{[k,n]}Q_{5}^{[k,n]}
=12\sum_k\sum_{n=1}^{P_2-1}Q_{1}^{[k,n]}Q_{5}^{[k,n]}
=\frac{3\pi^2}{4}\sum_k\alpha_k\mu_k\sum_{n=1}^{P_2-1}(2n+1)^2\nonumber\\[1mm]
&=\pi^2\sum_k\alpha_k\mu_k\left(P_2^3-\frac{P_2}{4}-\frac{3}{4}\right),
\label{eq:cqmTypeI}
\end{align}
where we have used the absolute values of the charges in \eqref{eq:chargesTypeI}. For long quivers, $P_1\gg1$, the terms in \eqref{eq:cholTypeI} involving $\beta_k$ and $\nu_k$ are subleading. For $P_2\gg1$ the term proportional to $u_0^2$ is subleading as well. Thus, at leading order both central charges then reduce to $\pi^2P_2^3\sum_k\alpha_k\mu_k$, which supports the slice-wise interpretation.

As anticipated above, the D3$'$ and D7$'$ branes enter the quantum mechanics through the electric large gauge transformations (see Section~\ref{sec:one}). To see this, note that in the $n$-th electric gauge patch \eqref{eq:DpelectricmagneticchargestypeI} gives
\begin{equation}
    Q^e_{1}=-n\,Q^m_{D7'},
    \qquad
    Q^e_{5}=-n\,Q^m_{D3'} .
\label{eq:baryonrelations}
\end{equation}
Their physical meaning follows from the Wess--Zumino couplings of the colour branes. The terms linear in the worldvolume gauge field contain schematically
\begin{equation}
    S^{WZ}_{Dp}
    \supset
    -\int \widehat F_p\wedge A ,
\end{equation}
and for the D1 and D5 stacks the relevant components are $\widehat F_{1,\rho}$ and $\widehat F_{5,\rho}$ respectively, which give $Q^m_{D7'}$ and $Q^m_{D3'}$. 
These couplings induce an electric tadpole for $A_t$ that must be cancelled by fundamental strings ending on the colour branes. Following \cite{Ramirez:2021tkd,Lozano:2020sae,Conti:2025djz}, we interpret each electric large gauge transformation as the creation of $|Q^m_{D7'}|$ such strings on the D1 stack and $|Q^m_{D3'}|$ on the D5 stack, as suggested by \eqref{eq:baryonrelations}. In the $n$-th electric gauge patch there are then $n|Q^m_{D7'}|$ F1 endpoints on the D1 stack and $n|Q^m_{D3'}|$ on the D5 stack, where on the second branch the patches are counted from $x_2=2\pi P_2$, the mirror image of $x_2=0$, as dictated by the reflection symmetry of the completion.

The resulting configurations,
\begin{equation}
    \mathrm{D1}\text{--}\mathrm{F1}\text{--}\mathrm{D7}'\,,
    \qquad
    \mathrm{D5}\text{--}\mathrm{F1}\text{--}\mathrm{D3}'\,,
\end{equation}
realise Wilson lines in completely antisymmetric representations and the associated baryon vertices of the two colour sectors, as in \cite{Ramirez:2021tkd,Lozano:2020sae,Conti:2025djz}. The D1--D7$'$ and D5--D3$'$ systems have eight mixed Neumann--Dirichlet directions, so the lowest modes of the strings stretched between them are fermions. At most one string can then stretch between a given pair of branes, and integrating out these fermions gives a Wilson line in a completely antisymmetric representation \cite{Yamaguchi:2006tq,Gomis:2006sb}.
The relations for the D3 and D7 branes have the same structure, but since these are flavour branes, no baryon vertices are associated with them. Besides, the D5$'$ and D1$'$ charges entering these relations are not independent, since they are induced on the D7$'$ and D3$'$ branes, see \eqref{eq:inducedD1D5}.
\subsection{Quantum mechanics of backgrounds beyond Type A}
\label{subsec:qmGeneral}

The Type A backgrounds and the double non-Abelian T-dual in Section~\ref{sec:FTtypeI} are particular members of the class \eqref{eq:NSNSNiallsol}--\eqref{eq:RRNiallsol}, and the results above shed light on how to interpret more general solutions. We consider separable configurations \eqref{eq:spliting} with $h_1$, $h_2$ as in \eqref{eq:hs} and $g_1$, $g_2$ as in \eqref{eq:gs}, whose slopes can change at arbitrary interfaces $x_1=\pi k$ and $x_2=\pi n$. Here we take $h_v$ and $g_v$ to be linear, so that $\dot g_v^{\,2}\equiv w_0^2$ is constant. This choice also covers the Type A profiles, for which $g_v$ is reflected at the turning point but $\dot g_v^{\,2}$ does not change.

The charges in the cell $[k,n]$ follow from \eqref{eq:RRNiallsol} as in Section~\ref{subsec:x2asanisometry}, with the same orientation. The colour charges come from integrating $\widehat F_7$ and $\widehat F_3$ over $x_2\in[\pi n,\pi(n+1)]$ at fixed $x_1$, whereas the flavour charges are obtained from the delta functions of $h_1''$ and $h_2''$ in \eqref{eq:BianchiFT}. One finds
\begin{align}
Q_{1}^{[k,n]} &= c_0\,\alpha_k\Big(\varrho_n+\frac{\gamma_n}{2}\Big)+\Delta Q_{1}^{[k,n]},
&
Q_{5}^{[k,n]} &= c_0\,\mu_k\Big(\sigma_n+\frac{\theta_n}{2}\Big)+\Delta Q_{5}^{[k,n]},
\nonumber\\[1mm]
Q_{3}^{[k,n]} &= c_0\,(\beta_{k-1}-\beta_k)\Big(\varrho_n+\frac{\gamma_n}{2}\Big),
&
Q_{7}^{[k,n]} &= c_0\,(\nu_{k-1}-\nu_k)\Big(\sigma_n+\frac{\theta_n}{2}\Big),
\label{eq:chargesGeneral}
\end{align}
where the contributions of $v$ to the colour charges are
\begin{align}
\Delta Q_{1}^{[k,n]}&
=-\frac{c_0}{\pi}\,h_v(\pi k)\,\frac{h_v}{h_2}\,\dot g_v\,\bigg[\frac{g_v}{g_2}\bigg]_{x_2=\pi n}^{x_2=\pi(n+1)},\nonumber\\[1mm]
\Delta Q_{5}^{[k,n]}&
=-\frac{c_0}{\pi}\,h_v(\pi k)\,\frac{h_v}{h_1}\,\dot g_v\,\bigg[\frac{g_v}{g_1}\bigg]_{x_2=\pi n}^{x_2=\pi(n+1)}.
\label{eq:DeltaQGeneral}
\end{align}
The first expressions follow from the exact terms in $\widehat F_7$ and $\widehat F_3$ and hold for any $v$. In the second ones we have used that $h_v$ and $g_v$ are linear.

The flavour charges and the first terms of the colour charges are those of \eqref{eq:charges_X2isometry}, with $\varrho$ and $\sigma$ replaced by the averages of $g_1$ and $g_2$ over the interval, $\varrho_n+\gamma_n/2$ and $\sigma_n+\theta_n/2$. The corrections \eqref{eq:DeltaQGeneral} vanish for $\dot g_v=0$, and $\Delta Q_1^{[k,n]}$ ($\Delta Q_5^{[k,n]}$) also vanishes when $g_v$ is proportional to $g_2$ ($g_1$) on the interval. This is the case for the Type A backgrounds, for which one recovers \eqref{eq:chargesTypeI}. Otherwise the corrections depend on $x_1$ through $h_v/h_2$ and $h_v/h_1$. When the profiles along $x_2$ grow linearly with $n$, the numerator above is at most of order $n$ and the denominator of order $n^2$, so the corrections decrease at least as $1/n$, while the first terms in \eqref{eq:chargesGeneral} grow as $n$. The charges are therefore well defined at leading order, where they reduce to these first terms.

For these profiles the integral in \eqref{eq:ccholfamily} factorises in each cell, and one finds
\begin{align}
c_{\rm hol}=6c_0^2\sum_{k,n}\bigg[&\Big(\alpha_k\mu_k+\frac{1}{2}(\alpha_k\nu_k+\beta_k\mu_k)+\frac{1}{3}\beta_k\nu_k\Big)\Big(\Big(\varrho_n+\frac{\gamma_n}{2}\Big)\Big(\sigma_n+\frac{\theta_n}{2}\Big)+\frac{\gamma_n\theta_n}{12}\Big)\nonumber\\
&-\frac{w_0^2}{3}\Big(h_v(\pi k)^2+h_v(\pi k)\,h_v(\pi(k+1))+h_v(\pi(k+1))^2\Big)\bigg]~,
\label{eq:cholGeneral}
\end{align}
The last term is the contribution of $\dot v^{\,2}=h_v^2w_0^2$, and it is the same for all the cells at fixed $k$. As in Section~\ref{subsec:qmTypeI}, we take the quantum mechanics to be a collection of copies of the quiver of Figure~\ref{fig:quiverX1}, one for each interval along $x_2$. Using the charges \eqref{eq:chargesGeneral}, its field theory central charge is
\begin{equation}
c_{qm}=6\sum_{k,n}Q_{1}^{[k,n]}Q_{5}^{[k,n]}=6c_0^2\sum_{k,n}\alpha_k\mu_k\Big(\varrho_n+\frac{\gamma_n}{2}\Big)\Big(\sigma_n+\frac{\theta_n}{2}\Big)+\delta c_{qm}~,
\label{eq:cqmGeneral}
\end{equation}
where $\delta c_{qm}$ collects the terms linear and quadratic in $\Delta Q_1^{[k,n]}$ and $\Delta Q_5^{[k,n]}$.

In general, the two central charges agree at leading order. The reason is that, cell by cell, both are fixed by $h_3$ and $h_7$. Apart from the term proportional to $w_0^2$, $c_{\rm hol}$ is $6c_0^2$ times the sum over cells of the average of $h_3h_7$ over each cell. Similarly, up to terms involving $\beta_k$ and $\nu_k$, the first terms of $Q_1^{[k,n]}$ and $Q_5^{[k,n]}$ are $c_0$ times the averages of $h_3$ and $h_7$. Therefore, the first term in \eqref{eq:cqmGeneral} is $6c_0^2$ times the sum of the products of these averages. The two expressions coincide when $h_3$ and $h_7$ vary slowly over each cell, that is, when $\alpha_k$, $\mu_k$, $\varrho_n$ and $\sigma_n$ are large compared with the slopes $\beta_k$, $\nu_k$, $\gamma_n$ and $\theta_n$. The differences arise from the terms involving $\beta_k$ and $\nu_k$ in \eqref{eq:cholGeneral}, which are subleading for long quivers along $x_1$ as in Section~\ref{subsec:qmUone}, and the term $\gamma_n\theta_n/12$, which is subleading when the profiles along $x_2$ are large compared with their slopes. The terms that depend on $v$ are also subleading. The contribution of each cell to the term proportional to $w_0^2$ does not depend on $n$, while the leading term grows with $\varrho_n\sigma_n$. Moreover, $\delta c_{qm}$ is suppressed by at least two powers of $n$ with respect to the first term in \eqref{eq:cqmGeneral}, as follows from the behaviour of $\Delta Q_{1,5}^{[k,n]}$. Finally, the geometry is only defined where $\Xi_1>0$, that is, where $h_1h_2g_1g_2>h_v^2w_0^2$. 
The agreement supports extending the description of Section~\ref{subsec:qmTypeI} to the whole family. In this limit, the dual quantum mechanics is a sum of copies of the quiver of Figure~\ref{fig:quiverX1}, one for each interval along $x_2$, with ranks given by the first terms in \eqref{eq:chargesGeneral}, as depicted in Figure~\ref{fig:quivercell}. This conclusion only involves the magnetic charges, so it does not depend on the baryon-vertex interpretation of \eqref{eq:baryonrelations}, which we discuss next.

The relations \eqref{eq:baryonrelations}, and hence the baryon-vertex interpretation, require further conditions. From \eqref{eq:RRNiallsol} one finds
%\begin{align}
%\widehat F_3\big|_{{\rm AdS}_2\wedge\mathrm{d}x_1}+n\pi\,\widehat F_1\big|_{\mathrm{d}x_1}&=c_0\Big[x_2\dot h_7-h_7+\partial_{x_1}\Big(\frac{(x_2\dot v-v)\,v'}{h_3}\Big)\Big],\nonumber\\
%\widehat F_7\big|_{{\rm AdS}_2\wedge{\rm CY}_2\wedge\mathrm{d}x_1}+n\pi\,\widehat F_5\big|_{{\rm CY}_2\wedge\mathrm{d}x_1}&=c_0\Big[h_3-x_2\dot h_3+\partial_{x_1}\Big(\frac{(v-x_2\dot v)\,v'}{h_7}\Big)\Big].
%\end{align}
%Once integrated over $x_1\in[\pi k,\pi(k+1)]$ at fixed $x_2$, the left-hand sides give $Q^e_{D1}+n\,Q^m_{D7'}$ and $Q^e_{D5}+n\,Q^m_{D3'}$, up to normalisation, and the right-hand sides give
\begin{align}
Q^e_{D1}+n\,Q^m_{D7'}&\;=c_0\Bigg[(\sigma_n-n\theta_n)\int_{\pi k}^{\pi(k+1)}h_2\,\mathrm{d}x_1+g_v(0)\,\frac{g_v}{g_1}\,\Big[\frac{h_vh_v'}{h_1}\Big]_{x_1=\pi k}^{x_1=\pi(k+1)}\Bigg],\nonumber\\
Q^e_{D5}+n\,Q^m_{D3'}&\;=c_0\Bigg[(\varrho_n-n\gamma_n)\int_{\pi k}^{\pi(k+1)}h_1\,\mathrm{d}x_1+g_v(0)\,\frac{g_v}{g_2}\,\Big[\frac{h_vh_v'}{h_2}\Big]_{x_1=\pi k}^{x_1=\pi(k+1)}\Bigg].
\label{eq:electricGeneral}
\end{align}
The first terms do not depend on $x_2$, while the second ones vary with $x_2$ as $g_v/g_{1,2}$. The relations \eqref{eq:baryonrelations} therefore hold if
\begin{itemize}
\item[(i)] $\varrho_n=n\gamma_n$ and $\sigma_n=n\theta_n$, that is, $g_1$ and $g_2$ are proportional to $x_2$ in the $n$-th interval, and
\item[(ii)] $g_v(0)=0$ or $h_v'=0$, that is, $g_v$ is also proportional to $x_2$, or $v$ does not depend on $x_1$.
\end{itemize}
Here $x_2=0$ is the point from which the electric patches are counted. If they are counted from $x_2=m\pi$, as on the second branch of the Type A completion, the conditions are imposed with respect to that point. Thus, a change of slope along $x_2$ is compatible with (i) only if the origin of the counting changes across it. The Type A backgrounds satisfy (i) and (ii), with $g_1$, $g_2$ and $g_v$ proportional to $x_2$ on the first branch and to $2\pi P_2-x_2$ on the second. This corresponds to the linear growth along the direction generated by the SL$(2,\mathbb{R})$ non-Abelian T-duality, so the baryon-vertex interpretation, in line with \cite{Conti:2025djz}, is tied to this choice of origin of the profiles along $x_2$. For more general profiles, the D3$'$ and D7$'$ branes at the interfaces along $x_2$ are still present, but their role in the quantum mechanics remains to be understood.

The case $\dot g_v=0$ is simpler. Then $\Xi_1=h_3h_7$, and there is no lower bound on $x_2$. The corrections \eqref{eq:DeltaQGeneral} vanish, the charges \eqref{eq:chargesGeneral} are exact and the term proportional to $w_0^2$ in \eqref{eq:cholGeneral} is absent. Since $g_v$ is then a non-vanishing constant, condition (ii) requires $h_v'=0$, that is, $v$ constant. In this case $\Xi_1=\Xi_2$, so the AdS$_2$ and S$^2$ factors have the same warp factor. Our proposal for the dual quantum mechanics of these backgrounds is the collection of quivers of Figure~\ref{fig:quiverX1}, one for each interval along $x_2$, with $\varrho$ and $\sigma$ in \eqref{eq:charges_X2isometry} replaced by the averages of $g_1$ and $g_2$ over that interval. Since there are no contributions from $v$, the two central charges differ only through the variation of the profiles inside each cell, and they agree in the holographic limit. The relations \eqref{eq:baryonrelations}, and hence the baryon vertices, hold whenever the profiles along $x_2$ satisfy (i).

\section{Conclusions}
\label{sec:conclusions}

In this paper we have used non-Abelian T-duality as a route to AdS$_2\times$S$^2$ solutions, since dualising an SL$(2,\mathbb{R})$ isometry of AdS$_3$ gives an AdS$_2$ factor and dualising an SU$(2)$ isometry of S$^3$ gives an S$^2$. As expected, the two orderings of the dualisations lead to the same solution, which can also be obtained in a single dualisation with respect to SL$(2,\mathbb{R})\times$SU$(2)$. The result does not depend on the seed solution, as long as the two groups act freely on it. For the massive Type IIA AdS$_3\times$S$^2\times$CY$_2\times$I solutions of \cite{Lozano:2019emq} considered here, the double dualisation leads to massless Type IIA. This is a general feature: only fluxes with all their legs along the six dualised directions, such as $F_6$ along AdS$_3\times$S$^3$, contribute to the dual Romans mass.

Our main focus has been the field theory interpretation of the Type IIB AdS$_2\times$S$^2\times$CY$_2\times\Sigma$ solutions constructed in \cite{Legramandi:2023fjr}. In the sector in which the backgrounds depend only on the coordinates $(x_1,x_2)$ of $\Sigma$, we have summarised their supergravity description, fluxes and charges, and read off the underlying brane intersection. We have shown that the Type A backgrounds belong to this class, and we have used this embedding to propose a global completion along the dual direction $x_2$, in which the solution is glued to its mirror image as in \cite{Ramirez:2021tkd}, with D3$'$ and D7$'$ branes at the turning point.

Most of our analysis has concerned the Type A backgrounds. Their brane configuration consists of D1 and D5 colour branes, NS5-branes and D3 and D7 flavour branes, together with the D3$'$ and D7$'$ branes at the turning point. The field theory analysis builds on the backgrounds with a U(1) isometry along $\partial_{x_2}$, whose $\mathcal{N}=4$ quantum mechanics was identified in \cite{Lozano:2020txg} by dimensional reduction of the two-dimensional $\mathcal{N}=(0,4)$ quivers dual to the parent AdS$_3$ solutions \cite{Lozano:2019zvg}. The quantum mechanics proposed here consists of one copy of this quiver for each interval along $x_2$. Each copy is the linear quiver along $x_1$ associated with the D1--D5--NS5 system and its D3 and D7 flavour branes. The resulting central charge agrees with the holographic one at leading order. The D3$'$ and D7$'$ branes do not give rise to flavour groups along $x_2$. They enter the quantum mechanics through fundamental strings stretched between them and the colour branes, which, as in \cite{Ramirez:2021tkd,Lozano:2020sae,Conti:2025djz}, we have interpreted as baryon vertices associated with Wilson lines in completely antisymmetric representations.

The double non-Abelian T-dual of AdS$_3\times$S$^3\times$CY$_2$, the near-horizon geometry of the D1--D5 system, also belongs to the Type A family, and therefore admits the same field theory description. For CY$_2=$T$^4$ this background arises, after an analytic continuation, as a limit of the $\lambda$-deformed SL$(2,\mathbb{R})\times$SU$(2)$ model \cite{Sfetsos:2014cea,Itsios:2023kma}, and its global completion may be a starting point for a field theory interpretation of the $\lambda$-deformed backgrounds in this limit.

What we have learnt from the Type A backgrounds has guided our interpretation of the class beyond Type A, for profiles that factorise in $x_1$ and $x_2$, with D3 and D7 branes distributed along $x_1$ and D3$'$ and D7$'$ branes along $x_2$. We have proposed that the dual quantum mechanics of these backgrounds is again a collection of copies of the same one-dimensional quiver, one for each interval along $x_2$, with ranks given by the D1 and D5 charges of the corresponding cell. The two central charges agree at leading order, since cell by cell both are fixed by the products of these charges. This proposal also covers the backgrounds in which the AdS$_2$ and S$^2$ factors have the same warp factor. The baryon-vertex interpretation, however, requires $g_1$ and $g_2$ to be proportional to $x_2$, measured from the point at which the electric gauge patches are counted, and either $g_v$ to be proportional to $x_2$ as well or $v$ to be independent of $x_1$. The Type A backgrounds satisfy these conditions, with the linear growth along $x_2$ generated by the SL$(2,\mathbb{R})$ dualisation. For more general profiles the role of the D3$'$ and D7$'$ branes in the quantum mechanics remains to be understood.

These results are first steps towards a field theory interpretation of the whole class of solutions of \cite{Legramandi:2023fjr}. It would be interesting to extend the analysis to the solutions of the class that do not factorise in $x_1$ and $x_2$. We hope to come back to this question in the near future.

\appendix

\section{Group theoretic conventions and single dualisation}
\label{app:groups}

\subsection{The Maurer-Cartan forms of $\mathrm{S^3}$}

A background containing $\mathrm{S^3}$ as a submanifold has a metric that exhibits an $\mathrm{SO(4)}$ isometry. The $\mathrm{SO(4)}$ group is locally isomorphic to $\mathrm{SU(2)_{L}} \times \mathrm{SU(2)_{R}}$, meaning that its Lie algebra is isomorphic to $\mathfrak{so}(4) \cong \mathfrak{su}(2) \oplus \mathfrak{su}(2)$. 

The radius-one $\mathrm{\, S^3}$ metric may be written in Euler angles as
\begin{equation}
    ds_{\, \mathrm{S^3}}^2 \= \frac{1}{4} \left( d\theta^2 + d\phi^2 + 2 \cos \theta \, d\phi \, d\psi + d\psi^2 \right)~. 
    \label{S3met}
\end{equation}
One can perform the dualisation with respect to one of the isometry groups, for instance the group $ \mathcal{G} = \mathrm{SU(2)_{L}}$. The generators of the algebra $\mathfrak{su}(2)$ are $t^{a} = \frac{\tau^{a}}{\sqrt{2}}$, where $\tau^{a}$ are the Pauli matrices with $\tau^{a} := \tau_{a}$ as a notational convention, then
\begin{equation}
\tau^1 \=  \begin{pmatrix} 0 & 1 \\ 1 & 0 \end{pmatrix}~, \qquad
\tau^2 \=  \begin{pmatrix} 0 & -i \\ i & 0 \end{pmatrix}~, \qquad
\tau^3 \=  \begin{pmatrix} 1 & 0 \\ 0 & -1 \end{pmatrix}~.
\end{equation}
The generators satisfy the following relations,
\begin{equation}
\label{eq:algebraSU(2)}
    \mathrm{Tr}\,(t^{a} \, t^{b}) \= \delta^{ab}, \quad [t^{a}, t^{b}] \= i \sqrt{2} \epsilon^{ab}_{\,\,\,\,\, c} \, t^{c}~.
\end{equation}
One can express a group element $g \in \mathrm{SU(2)}$ in the Euler parametrisation as 
\begin{equation}
 g \= e^{\frac{i}{2} \, \phi \,  \tau^{3}} \, e^{\frac{i}{2} \, \theta \, \tau^2} \, e^{\frac{i}{2} \, \psi \, \tau^{3}}~, \quad \text{where} \quad 0\leq \theta \leq \pi~, \quad 0\leq \psi < 4\pi~, \quad 0\leq \phi < 2\pi~.    
\end{equation}
Given this parametrisation, the left-invariant one-forms \eqref{MCforms} are written as
\begin{equation}
    \begin{split}
        L^1 & \= \cos\psi \, \sin \theta \, d\phi -\sin{\psi}\, d\theta~,\\
        L^2 & \= \cos{\psi}\, d\theta + \sin \psi \, \sin \theta \, d\phi~,\\
        L^3 & \= \cos \theta \, d\phi + d\psi~.
    \end{split}
\end{equation}

The corresponding group manifold is the smooth manifold underlying the Lie group $\mathrm{SU(2)}$ which is diffeomorphic to $\mathrm{S}^3$, and that is endowed with the Riemannian metric \eqref{S3met}. 

\subsection{The Maurer-Cartan forms of $\mathrm{AdS_{3}}$}

$\mathrm{AdS_{3}}$ is the group manifold $\mathcal{G} = \mathrm{SL(2,\mathbb{R})}$, which admits a global isometry group $\mathrm{SL(2,\mathbb{R})_{L}} \times \mathrm{SL(2,\mathbb{R})_{R}}$, acting on group elements $h \in \mathrm{SL}(2,\mathbb{R})$ via left and right multiplication:
\begin{equation}
    (g_{L},g_{R}) \cdot h \= g_{L}^{-1} \, h \, g_{R}~,
\end{equation}
where $g_{L} \in \mathrm{SL(2,\mathbb{R})_{L}}$ and $g_{R} \in \mathrm{SL(2,\mathbb{R})_{R}}$. For the purpose of performing the NATD transformation, we only consider one of these factors, namely $\mathrm{SL(2,\mathbb{R})_{L}}$\footnote{For simplicity, with a slight abuse of notation, we will call it $\mathrm{SL(2,\mathbb{R})}$ from now on.}, which is a non-compact group acting freely on the group manifold $\mathcal{G}$ via left multiplication.

The background fields obtained from \eqref{actionMC} are written in terms of the Maurer-Cartan-forms $L^{i}$, which derive from the left-invariant one-form $g^{-1} \mathrm{d}g$, where $g \in \mathcal{G}$. This form is invariant under left multiplication $g \mapsto g^{-1}_{L}\, g$. As a result, the geometry remains invariant under this transformation.
Although we are working with $\mathrm{SL(2,\mathbb{R})}$, it is convenient to express its elements using the Euler-angle parametrisation obtained via analytic continuation from $\mathrm{SU}(2)$ \footnote{This is motivated by the fact that both $\mathfrak{sl}(2,\mathbb{R})$ and $\mathfrak{su}(2)$ are real forms of the complex Lie algebra $\mathfrak{sl}(2,\mathbb{C})$.}. In this parametrisation, an element $\widetilde{g} \in \mathrm{SL}(2,\mathbb{R})$ can be written as
\begin{equation}
    \widetilde{g} \= e^{\frac{i}{2} \, \widetilde{\phi} \, \widetilde{\tau}^{3}} \, e^{\frac{i}{2} \, \widetilde{\theta} \, \widetilde{\tau}^2} \, e^{\frac{i}{2} \, \widetilde{\psi} \, \widetilde{\tau}^{3}}~, \quad \text{where} \quad 0\leq  \widetilde{\theta} \leq \pi~, \quad 0\leq  \widetilde{\psi} < \infty~, \quad 0\leq  \widetilde{\phi} < \infty~.
\end{equation}

The  generators $\widetilde{t}^{a}$ of $\mathfrak{sl}(2,\mathbb{R})$ can be obtained from those of $\mathfrak{su}(2)$ by analytic continuation, with $\widetilde{t}^{a}= \frac{\widetilde{\tau}^{a}}{\sqrt{2}}$, with $ \widetilde{\tau}^{a} := \widetilde{\tau}_{a}$, where 
\begin{equation}
\widetilde{\tau}^1 \= i\tau^{1} \= \begin{pmatrix} 0 & i \\ i & 0 \end{pmatrix}~, \qquad
\widetilde{\tau}^2 \= \tau^{2} \= \begin{pmatrix} 0 & -i \\ i & 0 \end{pmatrix}~, \qquad
\widetilde{\tau}^3 \= i \tau^{3} \= \begin{pmatrix} i & 0 \\ 0 & -i \end{pmatrix}~,
\end{equation}
such that $\widetilde{t}^{a}$ satisfy the algebra
\begin{equation}
\label{eq:algebraSl(2R)}
    \mathrm{Tr}(\,\widetilde{t}^{a} \, \widetilde{t}^{b} ) \= (-1)^{a} \, \delta^{ab}, \quad [\widetilde{t^{1}}, \widetilde{t^{2}}] \= i \sqrt{2} \, \widetilde{t^{3}}~, \quad[\widetilde{t^{2}}, \widetilde{t^{3}}] \= i \sqrt{2} \, \widetilde{t^{1}}, \quad [\widetilde{t^{3}}, \widetilde{t^{1}}] \= -i \sqrt{2} \, \widetilde{t^{2}}~,
\end{equation} 
where the trace is no longer positive definite, reflecting the Lorentzian signature of the group manifold.
With this parametrisation, the Maurer--Cartan one-forms read
\begin{equation}
    \begin{split}
        L^1 & \= \cosh{\widetilde{\psi}}\, d\widetilde{\theta} - \sinh \widetilde{\psi} \, \sin \widetilde{\theta} \, d\widetilde{\phi} ~,\\
         L^2 & \= \sinh{\widetilde{\psi}} \, d\widetilde{\theta} - \cosh \widetilde{\psi} \, \sin \widetilde{\theta} \, d \widetilde{\phi} ~,\\
        L^3 & \= - (\cos \widetilde{\theta} \, d \widetilde{\phi} + d \widetilde{\psi})~.
    \end{split}
\end{equation}

As a Lie group, $\mathrm{SL(2,\mathbb{R})}$ is also a smooth manifold which is diffeomorphic to $\mathrm{AdS_{3}}$, endowed with the Lorentzian metric
\begin{equation}
    ds_{\, \mathrm{AdS_3}}^2 \= \frac{1}{4} \left(- d \widetilde{\theta}^2 + d \widetilde{\phi}^2 + 2 \cos \widetilde{\theta} \, d \widetilde{\phi} \, d \widetilde{\psi} + d \widetilde{\psi}^2 \right)~,
    \label{AdS3met}
\end{equation}
which describes this space with unit radius of curvature.

\subsection{A single dualisation with respect to SL$(2,\mathbb{R})\times$SU$(2)$}
\label{sec:oneStep}

The two chains in Figure~\ref{Fig::Diagram} can be replaced by a single NATD with respect to the six-dimensional group $\mathcal{G}=\mathrm{SL}(2,\mathbb{R})\times\mathrm{SU}(2)$. This makes the commutativity of the two dualisations manifest and gives a direct route from AdS$_3\times$S$^3$ to AdS$_2\times$S$^2$.

We now let the indices $i,j$ in \eqref{actionMC} run over the six directions of $\mathcal{G}$, so that $\mu,\nu=1,\dots,4$ label the spectators, and split them as $i=(a,\alpha)$, with $a$ along SL$(2,\mathbb{R})$ and $\alpha$ along SU$(2)$. The Maurer--Cartan forms are $L^i=(\widetilde{L}^a,L^\alpha)$ and, since the two algebras commute, the only non-vanishing structure constants are $f^a{}_{bc}$ and $f^\alpha{}_{\beta\gamma}$. Gauging the full group, with $A_\pm$ and $v=(\widetilde{v}_a,v_\alpha)$ valued in $\mathfrak{sl}(2,\mathbb{R})\oplus\mathfrak{su}(2)$, the field strength splits into the two sectors and the procedure goes through as in Section~\ref{sec:NATD}, now with the $6\times6$ matrix
\begin{equation}
\mathcal{M}=\begin{pmatrix} \widetilde{M}_{ab} & E_{a\beta}\\ E_{\alpha b} & M_{\alpha\beta}\end{pmatrix}~,\qquad 
\widetilde{M}_{ab}=E_{ab}+f^{c}{}_{ab}\,\widetilde{v}_c~,\qquad M_{\alpha\beta}=E_{\alpha\beta}+f^{\gamma}{}_{\alpha\beta}\,v_\gamma~,
\label{eq:M6}
\end{equation}
notice that $a,b,c$ run over the 3 SL$(2,\mathbb{R})$ coordinates, and $\alpha, \beta, \gamma$ run over the 3 coordinates of SU$(2)$. The gauge choice $g=\mathbb{I}$ now fixes both group elements, and the six Lagrange multipliers become the dual coordinates. With the hyperbolic and spherical parametrisations of equations \eqref{eq:lagmultipliersSL} and \eqref{eq:lagmultipliersSU2} they give the $(r_1,\mathrm{AdS}_2)$ and $(r_2,\widetilde{\mathrm{S}}^2)$ sectors.

The off-diagonal blocks of \eqref{eq:M6} only contain the components of the background that mix the two groups. When $\widetilde{M}$ is invertible, the dilaton shift is controlled by
\begin{equation}
\det \mathcal{M}=\det\widetilde{M}\,\det M'~,\quad \text{where} \quad
M'_{\alpha\beta}=E_{\alpha\beta}-E_{\alpha a}(\widetilde{M}^{-1})^{ab}E_{b\beta}+f^{\gamma}{}_{\alpha\beta}\,v_\gamma~.
\label{eq:Schur}
\end{equation}
The matrix $M'$ is the one of the second dualisation in the chain that starts with SL$(2,\mathbb{R})$. Indeed, after dualising SL$(2,\mathbb{R})$ the SU$(2)$ directions play the role of spectators, and the Buscher rules shift their components to $E_{\alpha\beta}-E_{\alpha a}(\widetilde{M}^{-1})^{ab}E_{b\beta}$, which depend on the dual coordinates $\widetilde{v}_a$ through $\widetilde{M}$. Starting instead with SU$(2)$ gives the analogous factorisation with the roles of the two blocks exchanged. The same structure appears in $\mathcal{M}^{-1}$, so the NS sector obtained in one step coincides with the one obtained along either chain. This is simply the statement that integrating out the two gauge fields together or one after the other gives the same result, and it only uses that the two groups commute and act freely.

Whenever the background has no mixed components between the AdS$_3$ and S$^3$ directions, neither in the metric nor in $B_2$, one has $E_{a\beta}=0$. Then $M'_{\alpha\beta}=M_{\alpha\beta}$ and
\begin{equation}
\mathcal{M}=\widetilde{M}\oplus M~,\qquad \det \mathcal{M}=\det\widetilde{M}\,\det M~,
\end{equation}
so the dual metric and $B_2$ are the sum of the SL$(2,\mathbb{R})$ and SU$(2)$ contributions, and $e^{-2\widehat{\Phi}}$ is the product of the two dilaton factors. This holds for any solution of this form. The parent solution of Appendix~\ref{app:B-Appendix} is an example, and the factorisation is precisely what one sees in the Type C background, where $\Delta_1$ and $\Delta_2$ enter independently.

The RR sector behaves in the same way. The Lorentz transformation relating the two dual frames obtained in one step factorises as $\Lambda=\Lambda_{\mathrm{SU}(2)}\Lambda_{\mathrm{SL}(2,\mathbb{R})}$, where each factor is the one found in the corresponding step of either chain, so that, up to a sign,
\begin{equation}
\Omega=\Omega_{\mathrm{SU}(2)}\,\Omega_{\mathrm{SL}(2,\mathbb{R})}~.
\end{equation}
Each factor acts non-trivially on three frame directions and has $\det\Lambda=-1$, so it is odd in gamma matrices and $\Omega$ is even. The six-dimensional dualisation therefore preserves the chirality of the theory, mapping Type IIA into Type IIA and Type IIB into Type IIB. Since $\Omega$ only contains products of two, four and six gamma matrices along the dualised directions, the Romans mass of the original solution does not contribute to $\widehat{F}_0$, which can only receive contributions from fluxes with all their legs along these six directions. A solution of massive Type IIA supergravity without such flux components is therefore mapped into a solution of massless Type IIA. The Type C solution, obtained from the massive Type IIA parent, is an example of both. With $\widehat{P}=P\cdot\Omega^{-1}$, the two chains give the same RR sector up to an overall sign, which is a symmetry of the equations of motion. When $E_{a\beta}=0$, as for our parent solution, $\Lambda$ is block diagonal in the original frame and each factor is given explicitly by \eqref{eq.NATD-lorentz-spinor}. Supersymmetry is preserved when the Killing spinors are not charged under the dualised isometries, that is, when in the frame of left-invariant forms they do not depend on the coordinates of $\mathcal{G}$ \cite{Kelekci:2014ima}. For $\mathcal{G}=\mathrm{SL}(2,\mathbb{R})\times\mathrm{SU}(2)$ this amounts to the same condition for each factor separately, and it has to be checked for the particular background under consideration.

This construction is independent of the detailed form of the parent solution, provided that the required $\mathrm{SL(2,\mathbb{R})} \times \mathrm{SU(2)}$ isometries are present. Any background admitting these freely acting isometries associated with the AdS$_3$ and S$^3$ factors, and for which at least some of the Killing spinors are invariant, can be taken to a supersymmetric AdS$_2\times$S$^2$ solution in a single step, preserving the supersymmetries associated with those spinors.

\section{Parent geometry: small $\mathcal{N}=(0,4)$ AdS$_3\times$S$^2$ in massive IIA}
\label{app:B-Appendix}

In this appendix we summarise the AdS$_3\times$S$^2\times$M$_4\times$I solutions of massive Type IIA supergravity with small $\mathcal{N}=(0,4)$ supersymmetry constructed in \cite{Lozano:2019emq}, from which all the backgrounds of Section~\ref{sec:NATD} are obtained. We restrict to M$_4=$CY$_2$ and to backgrounds that do not depend on the coordinates of the CY$_2$, and refer to \cite{Lozano:2019emq,Lozano:2019jza,Lozano:2019zvg} for details and to \cite{Lozano:2021xxs} for a review. In string frame, the NSNS sector reads
\begin{align}
ds^2&=\frac{u}{\sqrt{h_4h_8}}\left(ds^2_{\mathrm{AdS}_3}+\frac{h_4h_8}{\Delta}\,ds^2_{\mathrm{S}^2}\right)+\sqrt{\frac{h_4}{h_8}}\,ds^2_{\mathrm{CY}_2}+\frac{\sqrt{h_4h_8}}{u}\,d\rho^2~,\nonumber\\[1mm]
e^{-2\Phi}&=\frac{h_8^{3/2}}{4\,u\,h_4^{1/2}}\,\Delta~,\qquad H_3=\frac{1}{2}\,d\left(-\rho+\frac{uu'}{\Delta}\right)\wedge\mathrm{vol}_{\mathrm{S}^2}~,\qquad \Delta=4h_4h_8+u'^2~,
\label{eq:parentNS}
\end{align}
where $h_4$, $h_8$ and $u$ depend only on $\rho$, which parametrises the interval, and primes denote derivatives with respect to $\rho$. The AdS$_3$ and the S$^2$ have unit radius. The RR fluxes are
\begin{align}
F_0&=h_8'~,\qquad F_2=-\frac{1}{2}\left(h_8-\frac{h_8'\,uu'}{\Delta}\right)\mathrm{vol}_{\mathrm{S}^2}~,\nonumber\\[1mm]
F_4&=\left(d\left(\frac{uu'}{2h_4}\right)+2h_8\,d\rho\right)\wedge\mathrm{vol}_{\mathrm{AdS}_3}-h_4'\,\mathrm{vol}_{\mathrm{CY}_2}~,
\label{eq:parentRR}
\end{align}
with $F_6=-\star_{10}F_4$, $F_8=\star_{10}F_2$ and $F_{10}=-\star_{10}F_0$. These are solutions of the equations of motion provided that, away from localised sources,
\begin{equation}
h_4''=0~,\qquad h_8''=0~,\qquad u''=0~.
\label{eq:parentConditions}
\end{equation}
The first two conditions follow from the Bianchi identities and the last one from supersymmetry. The solutions preserve eight supercharges, with the SU$(2)$ R-symmetry of the small $\mathcal{N}=(0,4)$ superconformal algebra realised on the S$^2$. 

For the SU$(2)$ branch of Section~\ref{sec:twoandthree} we take CY$_2=\mathbb{R}^4$, written as a cone over S$^3$,
\begin{equation}
ds^2_{\mathrm{CY}_2}=dr^2+r^2\,ds^2_{\mathrm{S}^3}~,\qquad \mathrm{vol}_{\mathrm{CY}_2}=r^3\,dr\wedge\mathrm{vol}_{\mathrm{S}^3}~,
\label{eq:parentR4}
\end{equation}
with $ds^2_{\mathrm{S}^3}$ the unit-radius metric \eqref{S3met}. The SU$(2)$ that we dualise acts freely on this S$^3$, and the functions $h_4$, $h_8$ and $u$ are invariant under this action.

Global solutions are obtained by taking $h_4$ and $h_8$ piecewise linear in $\rho$ \cite{Lozano:2019jza,Lozano:2019zvg},
\begin{eqnarray}
\begin{aligned}
h_4=\alpha_k+\dfrac{\beta_k}{2\pi}(\rho-2\pi k),
\quad\quad
h_8=\mu_k+\dfrac{\nu_k}{2\pi}(\rho-2\pi k)
\end{aligned}
\label{eq:parentProfiles}
\end{eqnarray}
these are linear functions for $\rho \in [k,k+1]$ with $\alpha_k=\sum_{j=0}^{k-1}\beta_j$ and $\mu_k=\sum_{j=0}^{k-1}\nu_j$. These profiles ensure that the metric and the dilaton are continuous. With $x_1=\rho/2$ these become the profiles \eqref{eq:hs}.

The magnetic component of $B_2$ requires a large gauge transformation $B_2\rightarrow B_2+\pi k\,\mathrm{vol}_{\mathrm{S}^2}$ each time a point $\rho=2\pi k$ is crossed, so there is one NS5-brane per interval. For a compact CY$_2$ with $\mathrm{Vol}(\mathrm{CY}_2)=16\pi^4$, the Page charges of the colour branes in the interval $[2\pi k,2\pi(k+1)]$ and the numbers of flavour branes at $\rho=2\pi k$ are
\begin{equation}
Q_{\mathrm{D}2}^{(k)}=\alpha_k~,\qquad Q_{\mathrm{D}6}^{(k)}=\mu_k~,\qquad \Delta Q_{\mathrm{D}4}^{(k)}=\beta_{k-1}-\beta_k~,\qquad \Delta Q_{\mathrm{D}8}^{(k)}=\nu_{k-1}-\nu_k~.
\label{eq:parentCharges}
\end{equation}
The resulting brane intersection is summarised in Table~\ref{table:parent}. The D2- and D6-branes are colour branes stretched between NS5-branes, while the D4- and D8-branes give rise to flavour groups. These are the brane set-ups of the two-dimensional $\mathcal{N}=(0,4)$ quivers of \cite{Lozano:2019jza,Lozano:2019zvg}, whose reduction gives the quantum mechanics of Section~\ref{subsec:qmUone}.

\begin{table}[h]
\centering
\renewcommand{\arraystretch}{1.2}
\begin{tabular}{c||cc|cccc|c|ccc}
\hline\hline
Brane & $x^0$ & $x^1$ & $x^2$ & $x^3$ & $x^4$ & $x^5$ & $x^6$ & $x^7$ & $x^8$ & $x^9$\\\hline\hline
D2  & $\bullet$ & $\bullet$ & - & - & - & - & $\bullet$ & - & - & - \\
D4  & $\bullet$ & $\bullet$ & - & - & - & - & - & $\bullet$ & $\bullet$ & $\bullet$ \\
D6  & $\bullet$ & $\bullet$ & $\bullet$ & $\bullet$ & $\bullet$ & $\bullet$ & $\bullet$ & - & - & - \\
D8  & $\bullet$ & $\bullet$ & $\bullet$ & $\bullet$ & $\bullet$ & $\bullet$ & - & $\bullet$ & $\bullet$ & $\bullet$ \\
NS5 & $\bullet$ & $\bullet$ & $\bullet$ & $\bullet$ & $\bullet$ & $\bullet$ & - & - & - & - \\
\hline\hline
\end{tabular}
\caption{Brane intersection underlying the parent solution \eqref{eq:parentNS}--\eqref{eq:parentRR}. The two-dimensional CFT lives along $(x^0,x^1)$, the directions $(x^2,\dots,x^5)$ span the CY$_2$, $x^6$ is the direction along $\rho$, and $(x^7,x^8,x^9)$ are the transverse directions to the D2- and D6- branes on which the SU$(2)$ R-symmetry acts.}
\label{table:parent}
\end{table}

\acknowledgments
We would like to thank Philip Candelas, Xenia de la Ossa, Pyry Kuusela, Yolanda Lozano, Niall Macpherson and Salomon Zacarias for very useful discussions. We are especially grateful to Yolanda Lozano, Niall Macpherson and Salomon Zacarias for their valuable comments on the draft. AR is supported by the Postdoctoral Fellowship Junior Marie Curie of the Research Foundation -- Flanders (FWO) grant 12ACZ25N, and by the FWO projects G003523N, G094523N and G0E2723N and the KU Leuven C1 project C16/25/01. AR was partially supported by the Simons Foundation through the award number 1023171-RC.
MMR has been supported by the Cluster of Excellence PRISMA+ (Project ID 390831469). MMR acknowledges the Sociedad Matemática Mexicana (SMM) together with the Fundación Sofía Kovalévskaia (SK) for their support through the Apoyo Sofía Kovalévskaia 2025, which contributed to the completion of this work. 

%\paragraph{Statement of AI use.}
%During the preparation of this paper we used Claude (Anthropic) as an assistant. Its main use was in the writing: it helped us to edit and rephrase the text, to check its internal consistency and to format the LaTeX source. We also used it to verify with computer algebra some of the analytic expressions and to work out some intermediate steps of these computations. The ideas, the constructions and their physical interpretation are those of the authors. %All the AI-assisted text and computations were checked by the authors, who take full responsibility for the content of the paper.

\newpage
\bibliography{Bibliography}

% Bibliography

%% [A] Recommended: using JHEP.bst file
%% \bibliographystyle{JHEP}
%% \bibliography{biblio.bib}

%% or
%% [B] Manual formatting (see below)
%% (i) We suggest to always provide author, title and journal data or doi:
%% in short all the informations that clearly identify a document.
%% (ii) please avoid comments such as "For a review'', "For some examples",
%% "and references therein" or move them in the text. In general, please leave only references in the bibliography and move all
%% accessory text in footnotes.
%% (iii) Also, please have only one work for each \bibitem.

%\begin{thebibliography}{99}

%\bibitem{a}
%Author,
%\emph{Title},
%\emph{J. Abbrev.} {\bf vol} (year) pg.

%\bibitem{b}
%Author,
%\emph{Title},
%arxiv:1234.5678.

%\bibitem{c}
%Author,
%\emph{Title},
%Publisher (year).

%\end{thebibliography}
\end{document}